\documentclass[12pt,a4paper,preprint,superscriptaddress]{revtex4-1}
\usepackage{float}
\usepackage{amsmath,amssymb}
\usepackage[utf8]{inputenc}
\usepackage{graphicx}
\usepackage[english]{babel}

\makeatletter\AtBeginDocument{\let\@elt\relax}\makeatother

\begin{document}

\title{Physics-separating artificial neural networks for predicting sputtering and thin film deposition of AlN in Ar/N$_2$ discharges on experimental timescales}
\date{\today}
\author{Tobias Gergs}
\email[]{tobias.gergs@rub.de}
\author{Thomas Mussenbrock}
\email[]{thomas.mussenbrock@rub.de}
\affiliation{Chair of Applied Electrodynamics and Plasma Technology, Department of Electrical Engineering and Information Science, Ruhr University Bochum, 44780 Bochum, Germany}
\author{Jan Trieschmann}
\email[]{jt@tf.uni-kiel.de}
\affiliation{Theoretical Electrical Engineering, Department of Electrical and Information Engineering, Kiel University, Kaiserstraße 2, 24143 Kiel, Germany}
\affiliation{Kiel Nano, Surface and Interface Science KiNSIS, Kiel University, Christian-Albrechts-Platz 4, 24118 Kiel, Germany}

\begin{abstract}
Understanding and modeling plasma-surface interactions frame a multi-scale as well as multi-physics problem. Scale-bridging machine learning surface surrogate models have been demonstrated to perceive the fundamental atomic fidelity for the physical vapor deposition of pure metals. However, the immense computational cost of the data-generating simulations render a practical application with predictions on relevant timescales impracticable. This issue is resolved in this work for the sputter deposition of AlN in Ar/N$_2$ discharges by developing a scheme that populates the parameter spaces effectively. Hybrid reactive molecular dynamics / time-stamped force-bias Monte Carlo simulations of randomized plasma-surface interactions / diffusion processes are used to setup a physics-separating artificial neural network. The application of this generic machine learning model to a specific experimental reference case study enables the systematic analysis of the particle flux emission as well as underlying system state (e.g., composition, mass density, stress, point defect structure) evolution within process times of up to 45 minutes.
\end{abstract}

\maketitle

\newpage


\section{Introduction}
\label{sec:introduction}

\begin{figure}[b]
\includegraphics[width=8cm]{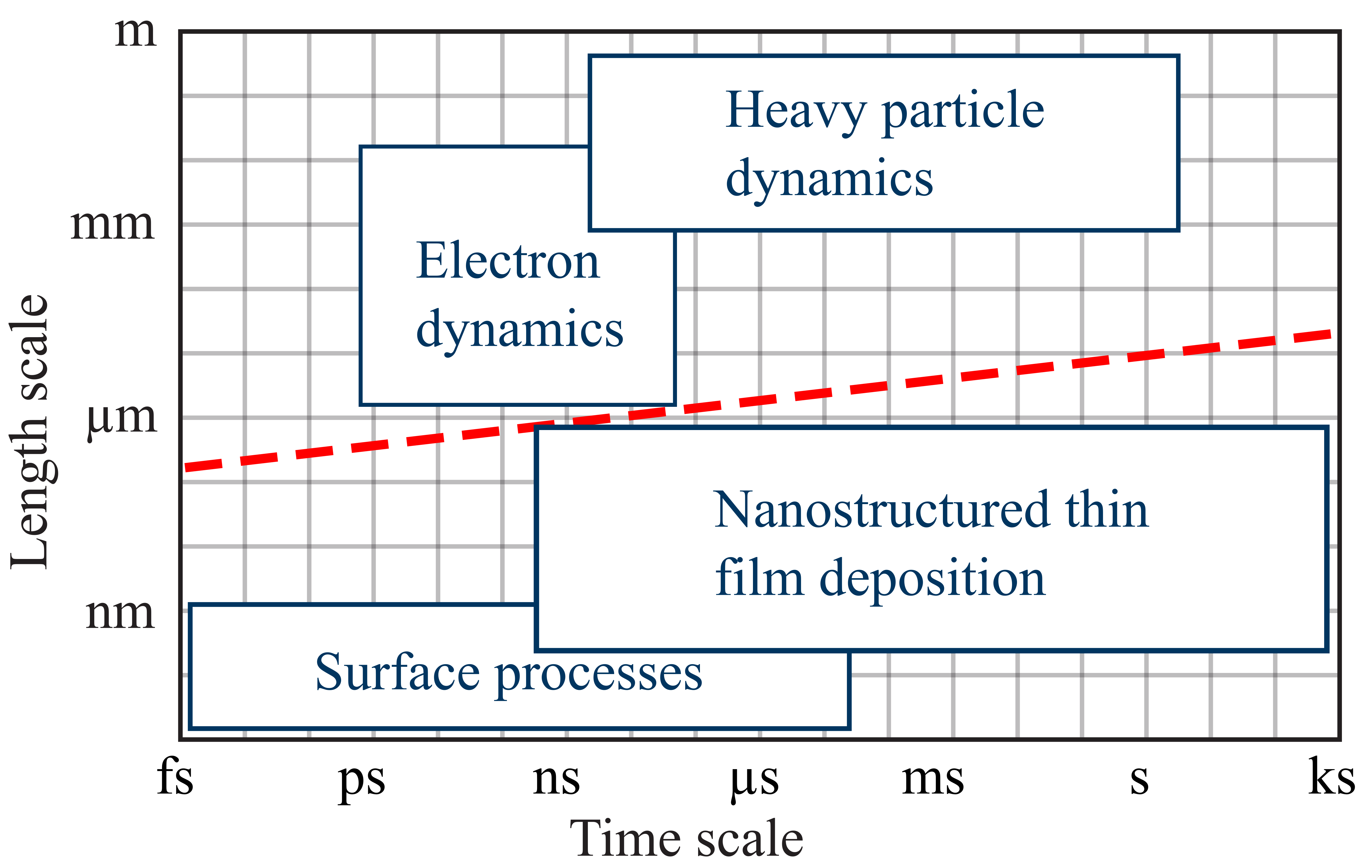}
\caption{Schematic of the physical time and length scales for thin film sputter depositions.}
\label{fig:Model_Scales_filled}
\end{figure}

In most technological applications of plasmas (e.g., thin film sputter deposition, catalysis) surfaces and, hence, plasma-surface interactions (e.g., growth, sputtering, surface chemical reactions) are involved \cite{kelly_magnetron_2000,gudmundsson_physics_2020,baptista_sputtering_2018,rossnagel_handbook_1990}. Analyzing, understanding, and modeling the last is considered to be essential for a knowledge-driven process design. However, the physics of these two states of matter (i.e., plasma, solid-state) demand for descriptions on length as well as time scales that differ in orders of magnitudes (see Figure~\ref{fig:Model_Scales_filled}) \cite{kruger_machine_2019, bird_molecular_1994, lieberman_principles_2005, callister_materials_2013}.

Common scale bridging solutions include event dependent coefficients, lookup-tables, and analytic formulas (e.g., Berg-model \cite{berg_fundamental_2005, depla_reactive_2008}, Sigmund--Thompson theory \cite{thompson_ii_1968, sigmund_theory_1969, sigmund_theory_1969-1}). However, they altogether lack a fundamental atomic fidelity.

An issue that has been addressed by applying machine learning (ML) models. They have been shown to be capable of describing physical processes relevant to plasma science with high accuracy while mitigating statistical noise, generalizing successfully \cite{kruger_machine_2019,preuss_bayesian_2019,ulissi_address_2017,diaw_multiscale_2020, kino_characterization_2021, adamovich_2022_2022, anirudh_2022_2022, gergs_efficient_2022,gergs_physics-separating_2022}. In particular, a series of ML plasma-surface interaction (PSI) surrogate models have been proposed for the sputter deposition of Ti$_ {1-x}$Al$_x$ thin films. First, a multi-layer-perceptron (MLP) was trained to predict the Ar$^+$ ion bombardment induced sputtering of a Ti$_{0.5}$Al$_{0.5}$ composite target \cite{kruger_machine_2019}. Second, a more advanced artificial neural network (ANN) combining a dedicated mapper network with the decoder of a $\beta$-variational autoencoder ($\beta$-VAE \cite{kingma_auto-encoding_2013, rezende_stochastic_2014, higgins_beta-vae_2016, burgess_understanding_2017, doersch_tutorial_2021}) was established for Ti$_ {1-x}$Al$_x$ composite targets \cite{gergs_efficient_2022}. Therein, the stoichiometry has been introduced as a basic surface state descriptor. Both studies are based on transport of ions in matter (TRIM) simulation data. Further, a physics-separating artificial neural network (PSNN) was proposed to describe the PSIs at the substrate as well as target in a generalized manner for Al and Ar as material system and working gas, respectively. The PSNN consists of two conditional variational autoencoders (CVAEs \cite{gergs_physics-separating_2022,sohn_learning_2015,doersch_tutorial_2021}). One describes the PSIs (e.g., sputtering, ion bombardment induced damage formation). The other one describes the conversion of the defect structure (i.e., ring statistical connectivity profile \cite{drabold2005models,cobb1996ab,zhang2000structural}) to the surface state (i.e., stoichiometry, mass density, biaxial stress, tensile stress). It was demonstrated that both (i.e., defect structure, surface state) are sufficient for a complete system description that may evolve in time. However, being based on molecular dynamics (MD) simulations for data generation, the latter was limited to the impingement of two consecutive particle doses (in total: $2.42\times10^{15}$ particles/cm$^2$) due to the immense computational cost. Hence, the input parameter space (i.e., particle flux composition, ion energy, surface state) was found to be sampled insufficiently to setup a long-term evolution ML PSI surrogate model for the sputter deposition of metal thin films.

In this work, the concept of a ML surface surrogate model is advanced by -- among other aspects -- proposing a randomized data generating scheme which enables PSNNs to predict the reactive sputter deposition of AlN thin films in Ar/N$_2$ discharges for up to hours. The considered process is relevant for the preparation of hard coatings, protective wear (e.g., transition metal aluminium nitride, transition metal aluminium oxynitride), and energy harvesting (scavenging) \citep{gibson_quantum_2018,iqbal_reactive_2018,
elfrink_vibration_2009,batra_piezoelectric_2016,beeby_energy_2006}. This manuscript is structured as follows: The considered scenario is presented in Section~\ref{sec:setup}. In Section~\ref{sec:methods}, applied methods and parameters are described. The results are presented and discussed in Section~\ref{sec:results}. Finally, conclusions are drawn in Section~\ref{sec:conclusion}.


\section{Setup}
\label{sec:setup}

The general scenario of an Ar/N$_2$ plasma discharge interacting with AlN surfaces is considered. While the gas discharge and sputtered particle transport dynamics are considered predetermined, the focus is on the substrate side AlN thin film deposition. The target side sputtering of AlN is not of main concern, but is included up to the maximum considered ion energy (i.e., 300\,eV). The key aspect for robust and reliable data-driven ML model development is to efficiently populate the parameter space relevant for representing the dynamics of PSI and diffusion. This is achieved by random sampling of a given number of initial AlN bulk systems, which are subsequently subject to a series of diffusion process and PSI simulations (e.g., ion bombardment). The corresponding evolution is recorded and used for ML. A brief description of the procedure is as follows:

\begin{figure}
\includegraphics[width=8cm]{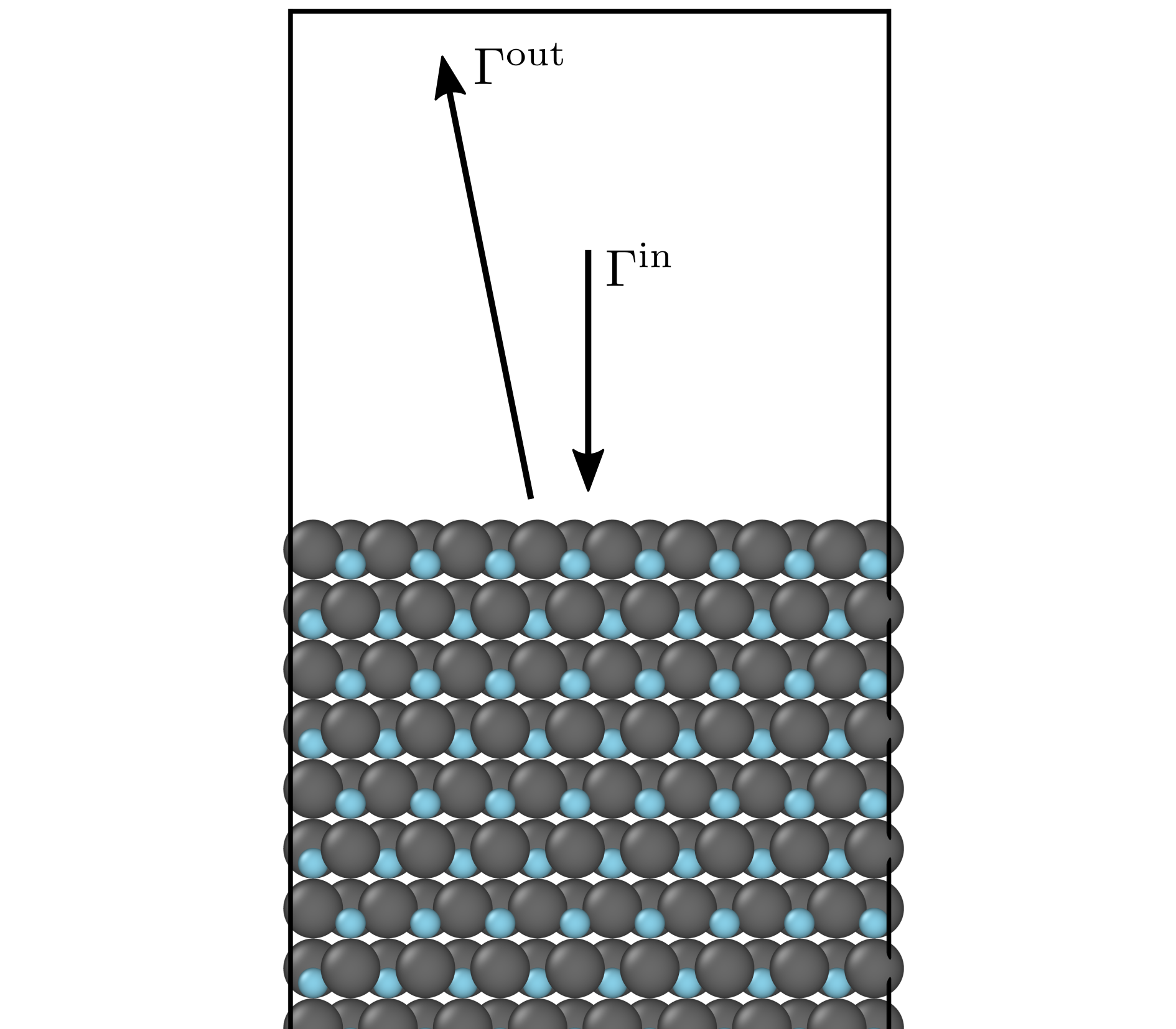}
\caption{Illustration of the PSI setup. The atom configuration is rendered with the Open Visualization Tool (OVITO) \cite{stukowski_visualization_2009}. Al and N atoms are colored gray and light blue, respectively.}
\label{fig:AtomGroups_motivation}
\end{figure}

\paragraph*{System state}
A bulk wurtzite AlN supercell is considered with a point defect structure that includes up to 5 \% Ar, 10 \% Al, and 10 \% N interstitials as well as 20 \% Al and 20 \% N vacancies. The defect structure is assumed to define the system sufficiently \cite{karimi_aghda_unravelling_2021,gergs_physics-separating_2022}. Complementing properties are determined after the atom configuration is relaxed. The system is characterized by the mass density $\rho$, lattice constant $a$, heat of formation $\Delta H_\mathrm{f}$, bulk modulus $B_0$, its derivative $B_0^\prime$, and 12 point defect populations $\rho_\mathrm{v_{Al}}$, $\rho_\mathrm{Al_N}$, $\rho_\mathrm{Al_i}$, $\rho_\mathrm{v_N}$, $\rho_\mathrm{N_{Al}}$, $\rho_\mathrm{(N\text{-}N)_{Al}}$, $\rho_\mathrm{N_i}$, $\rho_\mathrm{(N\text{-}N)_N}$, $\rho_\mathrm{(N\text{-}N)_i}$, $\rho_\mathrm{Ar_{Al}}$, $\rho_\mathrm{Ar_N}$, $\rho_\mathrm{Ar_i}$. The Kröger-Vink notation is used for the defect types (subscripts) \cite{KROGER1956307}. The defect populations define the total number of atoms in the system:
\begin{equation}
n_\mathrm{tot} = ({1+\rho_\mathrm{v_{Al}}+\rho_\mathrm{v_{N}}-\rho_\mathrm{{Al}_{i}}-\rho_\mathrm{{N}_{i}}-\rho_\mathrm{{Ar}_{i}}-2\rho_\mathrm{{(N\text{-}N)}_{i}}-\rho_\mathrm{{(N\text{-}N)}_{N}}-\rho_\mathrm{{(N\text{-}N)}_{Al}}})^{-1}{n_\mathrm{tot}^\mathrm{ideal}}
\label{eq:NGes}
\end{equation}
$n_\mathrm{tot}^\mathrm{ideal}$ refers to the total number of atoms in the ideal AlN supercell (8 atoms per unit cell). The point defect structure defines the Al, N, and Ar concentrations $c_\mathrm{Al}$, $c_\mathrm{N}$, and $c_\mathrm{Ar}$, which are denoted as the composition:
\begin{subequations}
\begin{align}
c_\mathrm{Al}&= 0.5\frac{n_\mathrm{tot}^\mathrm{ideal}}{n_\mathrm{tot}}-\rho_\mathrm{v_{Al}}+\rho_\mathrm{{Al}_i}+\rho_\mathrm{{Al}_N}-\rho_\mathrm{{N}_{Al}}-\rho_\mathrm{{(N\text{-}N)}_{Al}}-\rho_\mathrm{{Ar}_{Al}}
\label{eq:xAl}\\
c_\mathrm{N}&= 0.5\frac{n_\mathrm{tot}^\mathrm{ideal}}{n_\mathrm{tot}}-\rho_\mathrm{v_{N}}+\rho_\mathrm{N_{i}}+2\rho_\mathrm{{(N\text{-}N)}_{i}}+\rho_\mathrm{{(N\text{-}N)}_{N}}+2\rho_\mathrm{{(N\text{-}N)}_{Al}}-\rho_\mathrm{{Al}_N}-\rho_\mathrm{{Ar}_N}
\label{eq:xN}\\
c_\mathrm{Ar}&= \rho_\mathrm{{Ar}_i}+\rho_\mathrm{{Ar}_{Al}}+\rho_\mathrm{{Ar}_N}
\label{eq:xAr}
\end{align}
\end{subequations}
The first terms on the right hand side of Eqs.~\eqref{eq:xAl} and \eqref{eq:xN} refer to the ideal configuration, as $0.5 n_\mathrm{tot}^\mathrm{ideal}$ is the number of Al or N atoms when point defects are absent. The mass density is determined by the lattice constants, the total number of atoms $n_\mathrm{tot}$, and the composition:
\begin{equation}
\rho = \frac{m_\mathrm{Al} c_\mathrm{Al} + m_\mathrm{N} c_\mathrm{N} + m_\mathrm{Ar} c_\mathrm{Ar}}{\sqrt{3} n_\mathrm{uc}a^2 c} n_\mathrm{tot}
\label{eq:massdensity}
\end{equation}
$m_\mathrm{Al}$, $m_\mathrm{N}$, and $m_\mathrm{Ar}$ are the masses of Al, N, and Ar atoms, respectively. $n_\mathrm{uc}$ is the number of unit cells (detailed later) and the lattice constant $c=1.6a$ is kept constant (anisotropic deformations are suppresesed).

\paragraph*{Plasma-Surface Interaction and Diffusion}
For each initialized system, seven diffusion and PSI simulations are performed alternately (detailed later). First, the effect of bulk diffusion processes on the system state is studied. For this a temperature $T$ is imposed. Second, an AlN surface is obtained by cleaving the bulk system either in [100] or [002] direction. Third, the effect of individual particles $s$ (i.e., Al, N, N$_2$, Ar) bombarding the AlN surface with specified kinetic energies $E_\mathrm{kin}$ is investigated. The contribution from the plasma onto the surface is characterized by the particle fluxes $\Gamma^\mathrm{in}_s$, the kinetic energy of the particles $E_\mathrm{kin}$, and the species $s$. The emitted fluxes are denoted by $\Gamma^\mathrm{out}_s$.

The first and the last are used to setup two individual machine learning regression models (i.e., PSI-CVAE, Diffusion-CVAE) that eventually are used to form a PSNN.


\section{Methods}
\label{sec:methods}

First, the data generating hybrid reactive molecular dynamics (RMD) / time-stamped force-bias Monte Carlo (tfMC) simulations are described. Second, the data processing, training workflow and included metric are introduced. Third, the structure and information flow of the PSNN is outlined. Fourth, physics-constraints and their implementation are introduced. Fifth, the hyperparameter (HP) optimization is descried. Sixth, the production run is presented. 

\subsection{Hybrid reactive molecular dynamics / time-stamped force-bias Monte Carlo}
\label{ssec:RMD}

RMD, tfMC, and hybrid RMD/tfMC simulations are performed with the open-source Large-scale Atomic/Molecular Massively Parallel Simulator (LAMMPS) \cite{bal_time_2014,mees_uniform-acceptance_2012,neyts_combining_2014,plimpton_fast_1995,thompson_lammps_2022}. The interactions of AlN complexes are described by the third-generation charge-optimized many-body (COMB3) potential that is tapered with the Ziegler-Biersack-Littmark (ZBL) potential (COMB3/ZBL potential) to account for high-energy collisions by including screened nuclear repulsions \cite{liang_classical_2013,ziegler_stopping_1985,gergs_charge-optimized_2022}. The COMB3 formalism is outlined in \cite{liang_classical_2013}. The COMB3 AlN parameterization and combination with the ZBL potential is described in \cite{gergs_charge-optimized_2022}. Its predecessor was setup for nanostructures as well as heterogeneous interfaces and revisited to describe plasma-surface interactions more accurately (e.g., ion bombardment induced damage production) \cite{choudhary_dynamical_2016,gergs_charge-optimized_2022}. The atomic charges are equilibrated by applying the charge transfer equilibration (QTE$^+$) method to account for meaningful charge exchange during PSIs (e.g., ion bombardment, sputtering) \cite{gergs_generalized_2021}. In the following, charge equilibration refers to the application of the QTE$^+$ method with a timestep of 10$^{-2}$ fs. The exponents of the 1s Slater type orbitals used for the overlap integral computations are 0.668 \r A$^{-1}$ and 1.239 \r A$^{-1}$ for Al and N, respectively \cite{gergs_generalized_2021}.

\begin{figure*}
\includegraphics[width=16cm]{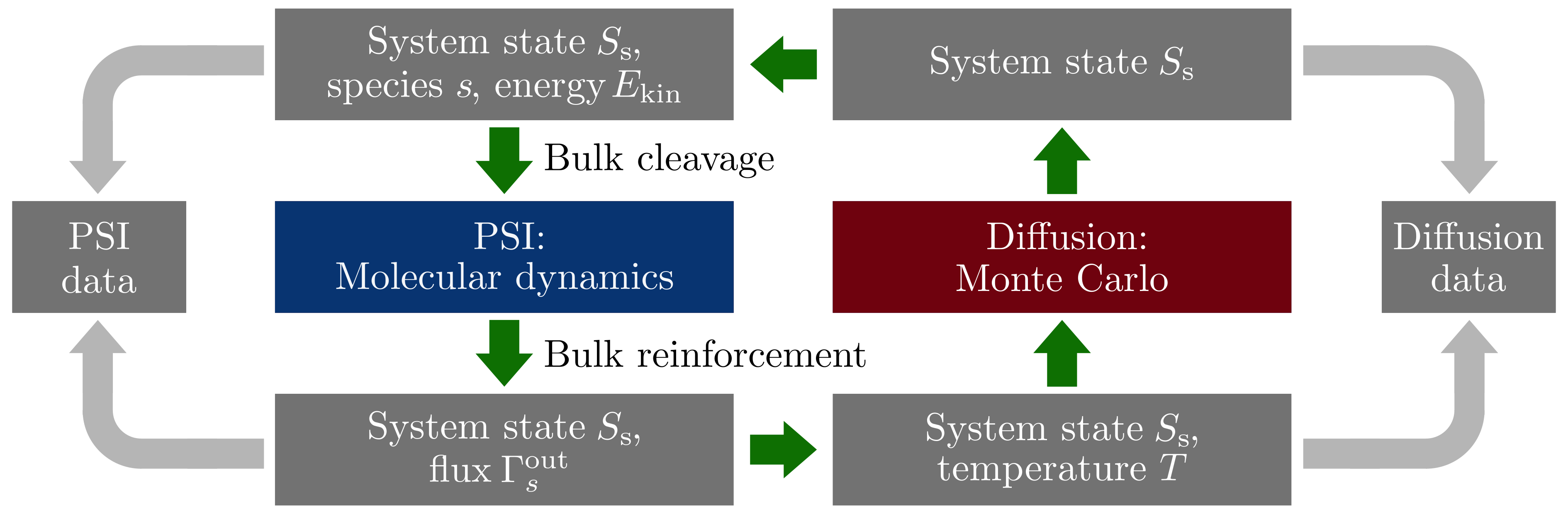}
\caption{Schematic of the workflow and information flow for the data generating hybrid RMD/tfMC simulations.}
\label{fig:DataGeneration}
\end{figure*}

\paragraph{System state initialization}
It has been argued and demonstrated that the defect structure is sufficient to describe a system \cite{karimi_aghda_unravelling_2021,gergs_physics-separating_2022}. Hence, the initial atom configuration is constructed by specifying the point defect structure. The Ar and Al (N) interstitial population $\rho_\mathrm{Ar_i}$ and $\rho_\mathrm{Al_i}$ are sampled from a normal distribution $\mathcal{N}(0,\sigma)$ with the standard deviations $3\sigma=5$ \% and $3\sigma=10$ \%, respectively. The N interstitial populations account for single as well as split interstitials (N-N) \cite{szallas_characterization_2014}. They are distinguished from each other at the end of the surface state initialization. The Al and N vacancy population $\rho_\mathrm{v_{Al}}$ and $\rho_\mathrm{v_N}$ are sampled from a normal distribution $\mathcal{N}(0,\sigma)$ with a standard deviation $3\sigma=20$ \%. Initially, no anti-sites (i.e., N$_\mathrm{Al}$, (N-N)$_\mathrm{Al}$, Al$_\mathrm{N}$, Ar$_\mathrm{Al}$, Ar$_\mathrm{N}$) are defined.

The surface orientation (i.e., AlN(002), AlN(100)) is determined by a coin flip. In either case, a bulk supercell consisting of $8\times5\times7$ orthorhombic unit cells is constructed with the lattice constants $a$=3.136 \r A and $c=1.6a$. The total number of atoms in the ideal AlN supercell (8 atoms per unit cell) is $n_\mathrm{tot}^\mathrm{ideal}=2240$. The targeted total number of atoms $n_\mathrm{tot}$ is calculated as a function of the point defect population following Eq.~\eqref{eq:NGes}. The absolute number of point defects is obtained by multiplying the total number of atoms with the individual point defect population.

First, Al and N vacancies are created by removing the required number of Al and N atoms from the system. Second, interstitials are taken care of by randomly inserting new atoms (i.e., Al, N, Ar) into the simulation domain. The Ar atoms' coordinate in surface normal direction is constrained to fall in between 5 \r A above and below the lower and upper boundary of the simulation domain, respectively. If the new atoms overlap with each other or old atoms, they are deleted. This second step is repeated until the correct number of Al, N and Ar atoms are generated. 

\begin{figure}
\includegraphics[width=8cm]{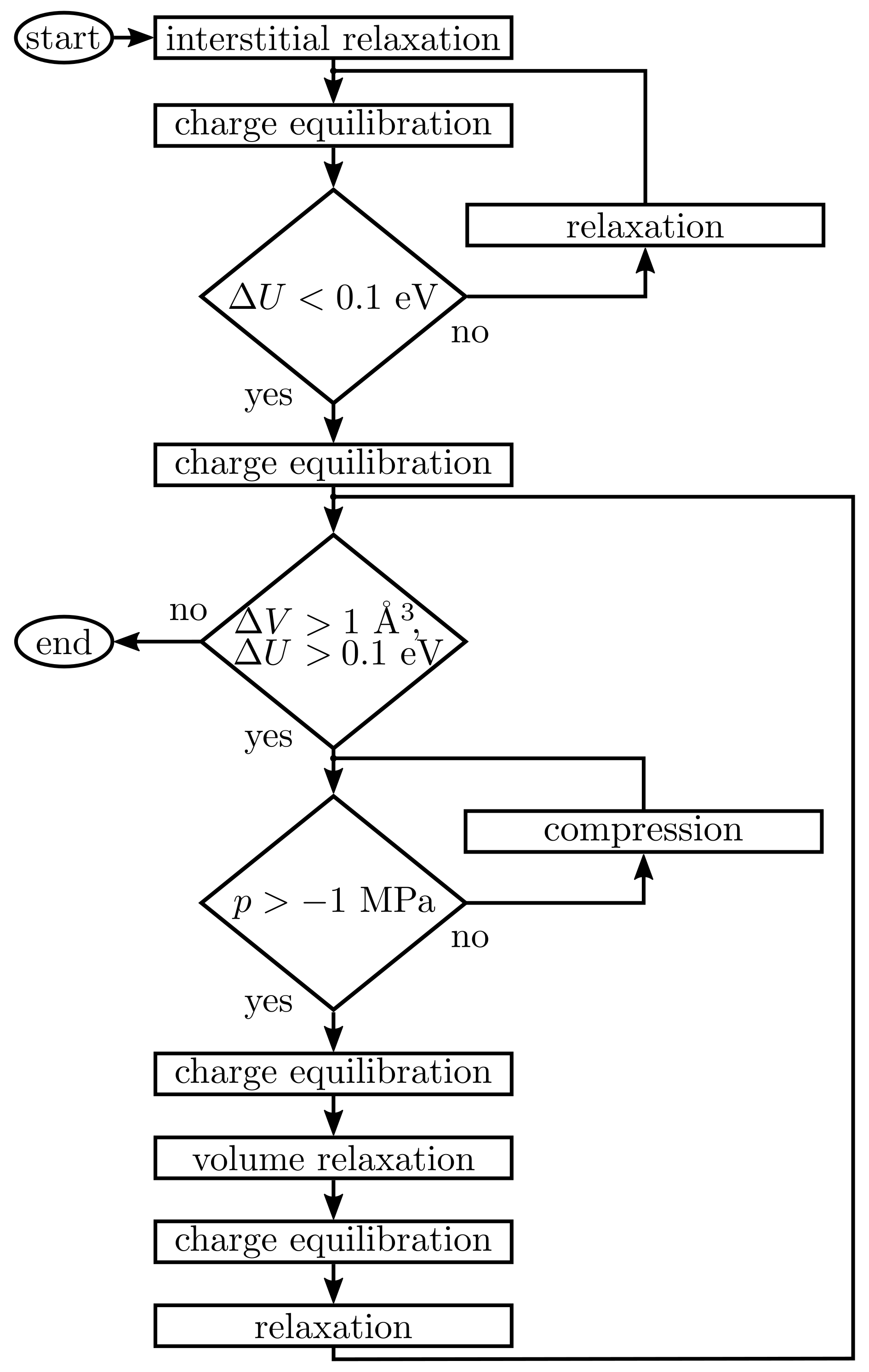}
\caption{Workflow of the bulk relaxation. Relaxation: Application of the conjugate gradient descent algorithm implemented in LAMMPS. The tolerance for the residual force on any atom is 1 eV/\r A. Charge equilibration: Performing one time step while the QTE$^+$ method is applied. $\Delta U$, $\Delta V$ and $p$ refer to the change of the potential energy, volume, and instantaneous pressure value, respectively. Compression: Application of the strain $-10^{-6}$ along each direction. Volume relaxation: In addition to the atom site relaxation, the simulation box dimensions are adjusted isotropically to remove the residual stress from the system.}
\label{fig:bulkOpt}
\end{figure}

The atom configuration is then relaxed. Minor discontinuities of the COMB3 interaction potential hinder the successful application of a single conjugate gradient descent algorithm. This issue is addressed by performing multiple energy minimizations (relaxations) as depicted in Figure~\ref{fig:bulkOpt}. The alternation between charge equilibration (i.e., applying QTE$^+$) and relaxation is meant to increase the computational efficiency. It is easier to relax an expanding than shrinking atom configuration. Hence, the system is compressed when the instantaneous pressure falls below -1 MPa. 

The resultant point defect structure is determined by comparing the position of each atom mapped into the unit cell with the Al as well as the N atom sites of the ideal AlN(002) or AlN(100) structures (periodic images are taken into account). The distance tolerance is defined by the halved Al-N bond length 1.9/2 \r A. Nitrogen split interstitials (N-N)$_\mathrm{i}$, (N-N)$_\mathrm{N}$ or anti-sites (N-N)$_\mathrm{Al}$ are identified by searching for interatomic distances between N atoms that fall below 1.5 \r A (the N-N bond length equals 1.3 \r A \cite{szallas_characterization_2014}). The number of Al and N vacancies are computed at last to fulfill the particle balances:
\begin{subequations}
\begin{align}
n_\mathrm{v_{Al}}&=0.5n_\mathrm{tot}^\mathrm{ideal}-n_\mathrm{tot,Al}+n_\mathrm{Al_{i}}+n_\mathrm{Al_{N}}-n_\mathrm{N_{Al}}-n_\mathrm{(N-N)_{Al}}-n_\mathrm{Ar_{Al}}
 \label{eq:VAlbalance}\\
n_\mathrm{v_{N}}&=0.5n_\mathrm{tot}^\mathrm{ideal}-n_\mathrm{tot,N}+n_\mathrm{N_{i}}+n_\mathrm{(N-N)_{N}}+2n_\mathrm{(N-N)_{i}}+n_\mathrm{N_{Al}}+2n_\mathrm{(N-N)_{Al}}-n_\mathrm{Al_{N}}-n_\mathrm{Ar_{N}}
 \label{eq:VNbalance}
\end{align}
\end{subequations}
The symbol $n$ describes the absolute number of point defects, while the indexes denote the particular point defect type. When the provided distance tolerance results in negative numbers for vacancies, Frenkel pairs (i.e., vacancies plus interstitials) are added to even out this diagnostic artifact. However, this procedure is applied rarely and is only meant to guarantee physically meaningful results (i.e., non-negative numbers of vacancies).

At last, the minimum of the potential energy and corresponding lattice constant is obtained by fitting the third-order Birch-Murnaghan equation of state (EOS) to the $p$-$V/n_\mathrm{tot}$ and $U/n_\mathrm{tot}$-$V/n_\mathrm{tot}$-curve of the just relaxed structure \cite{birch_finite_1947,murnaghan_compressibility_1944}. The system dimensions are scaled isotropically to evaluate ten strains distributed equidistantly in between $-10^{-2}$ and $10^{-2}$. The system is compressed before it is expanded. The atom sites are relaxed for each probed atom configuration.

\paragraph{Diffusion}
The tfMC method is applied for the simulation of the diffusion processes \cite{bal_time_2014,mees_uniform-acceptance_2012,neyts_combining_2014}. The maximal displacement length of the lightest atom (i.e., N) is $\Delta=0.19$\r A, that is approximately 10 \% of the typical nearest neighbor distance for AlN. The temperature is sampled from a uniform distribution in between 300 and 1000 K. Simultaneously, the QTE$^+$ method is applied with a time step of 10$^{-2}$ fs, tolerance of 1 V, and damping constant of 0.45 \cite{gergs_generalized_2021}. The simulation is run for $10^4$ steps. The resultant atom configuration is relaxed and evaluated as detailed previously (the interstitial relaxation is skipped).

\paragraph{Plasma-surface interaction}
The surface is established by cleaving the bulk system either in [100] or [002] direction and elongating the simulation domain in surface normal direction by additional 35 \r A. This value is defined to be the sum of 11 \r A and 24 \r A, that are attributed to the COMB3 cutoff radii and serve as a buffer for recognizing reflected or sputtered particles, respectively. The atom sites are adjusted by alternating between charge equilibration (i.e., QTE$^+$) for fixed atom configuraiton and relaxation (i.e., performing a conjugate gradient descent minimization until the residual force perceived by each atom falls below 1 eV/\r A) for a fixed charge distribution until the potential energy is changed by less than 0.1 eV per iteration. The surface slab is displaced randomly in both surface parallel directions to include different impingement sites (the impinging particles are centered above the surface slab).

\begin{figure}
\includegraphics[width=8cm]{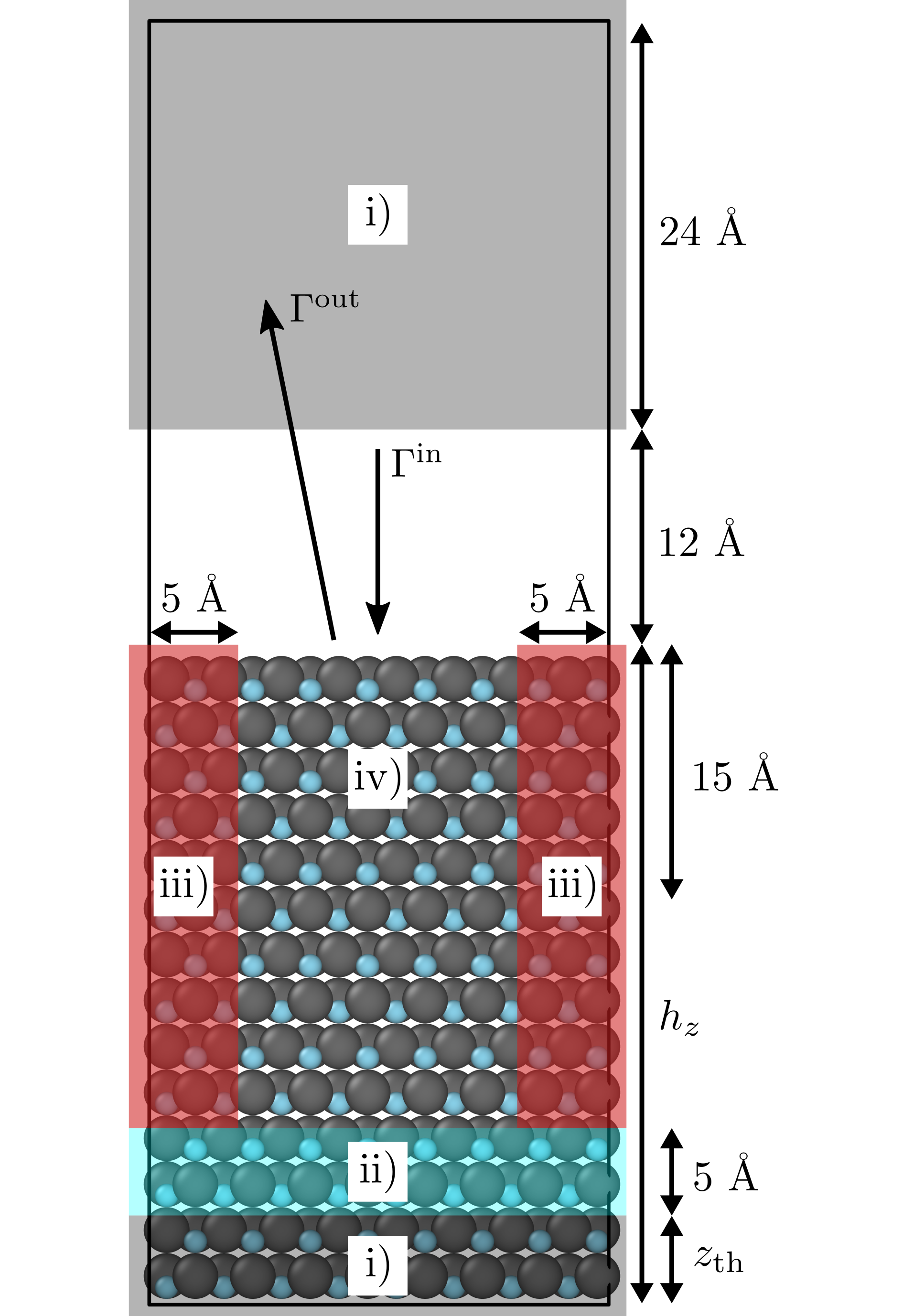}
\caption{Illustration of the PSI setup. The atom configuration is rendered with OVITO \cite{stukowski_visualization_2009}. Al and N atoms are colored gray and light blue, respectively. The regions containing i) excluded, ii) immobile, iii) temperature-controlled, and iv) all remaining atoms are colored transparent gray, light blue, red, and not at all respectively.}
\label{fig:AtomGroups}
\end{figure}

Following this procedure, the system can be subdivided into four regions as depicted in Figure~\ref{fig:AtomGroups}: i) Excluded atoms whose surface normal coordinate falls below a threshold $z_\mathrm{th}=h_z-15~\text{\r A}-(h_z-15~\text{\r A})E_\mathrm{kin}/E_\mathrm{kin,max}$. $z_\mathrm{th}$ is decreased linearly for increasing kinetic energies of impinging particles $E_\mathrm{kin}$, ranging from 0 eV to 300 eV ($E_\mathrm{kin,max}=300$ eV). $h_z$ is the surface slab height. Excluded atoms are not allowed to interact with any other atom, effectively reducing the surface slab thickness to reduce the computational cost. These atoms are also excluded from the charge equilibration. The interactions of reflected or sputtered particles with other atoms are excluded too, when their surface normal coordinate exceeds $h_z+12$ \r A. However, they are kept in the simulation domain to evaluate them later. The members of this group are updated dynamically (i.e., every 2000 steps). ii) Immobile atoms whose surface normal coordinate falls below a threshold $z_\mathrm{th}+5$ \r A. These are not evolved in time to anchor the surface slab in the simulation domain. iii) Mobile atoms whose distance to the left or right periodic boundary in the surface parallel directions falls below 5 \r A are coupled to a Langevin thermostat with a damping constant of 100 fs to gradually remove their kinetic energy, targeting 0 K. Ar atoms are always excluded. iv) All remaining atoms.

An impinging particle is created 11 \r A above the surface and centered laterally. Its species (i.e., Al, N, N$_2$, Ar) is determined randomly, whereas projectiles are assumed to be charge neutral prior interaction. The likelihood for each candidate is distributed equally among them. Its kinetic energy ranging from $0~\mathrm{eV}$ to $300~\mathrm{eV}$ is found by squaring a sample from a uniform distribution $\mathcal{U}(0~\sqrt{\mathrm{eV}},\sqrt{300~\mathrm{eV}})$. The atoms are assumed to hit the surface perpendicularly (the surface parallel components of its velocity vector equals zero). The time step equals 0.25 fs and is eventually lowered to secure that the maximum displacement and change in kinetic energy of any atom does not exceed 0.1 \r A and 0.01 eV, respectively. The simulation is run for 1 ps repeatedly until the temperature of the mobile atoms falls below 100 K. Impinging atoms that bypass the lower simulation domain boundary due to channeling are reinserted at a random position in surface normal direction (lateral coordinates are maintained) within the surface slab (overlapping with another atom by less than 0.5 \r A leads to a repetition).

To again transfer from a surface to a bulk configuration, the atom sites are relaxed as outlined previously. The random shifts in surface parallel directions outlined in the beginning of this section are reversed. The change of the surface normal coordinate of the uppermost temperature controlled atom $\Delta z_\mathrm{up}$ is used as a reference to invert the particle impingement induced thermal expansion of the mobile surface slab. The surface normal coordinates $z$ of all mobile atoms are updated by $z \rightarrow z-\Delta z_\mathrm{up}(z-z_\mathrm{th}-5~\text{\r A})/(z_\mathrm{up}-z_\mathrm{th}-5~\text{\r A})$ (assuming a linear expansion). All atoms that exceed the original surface slab height prior to the particle impingement are removed from the system. This includes reflected particles, sputtered particles, and in general atoms atop the surface (e.g., adatoms). The last are assumed to contribute to the film growth of following layers, but do not effect the subsurface region. Hence, they are neglected when making a prediction for the bulk system by reestablishing a periodic boundary in surface normal direction. This procedure has been validated by comparing lattice constants and stresses obtained with density functional theory based molecular dynamics thermal spike simulations to experimentally measured reference values for metal aluminium nitrides \cite{karimi_aghda_unravelling_2021,music_correlative_2017,holzapfel_influence_2022}. The resultant atom configuration is relaxed and further evaluated as detailed previously (the interstitial relaxation is omitted).

\subsection{Data preparation, training and metrics}
\label{ssec:data}

\paragraph{Data set splitting}
The data sets consisting of 6496 diffusion processes and 4470 PSIs are shuffled and split for the HP optimization to train, validate and test the ML model with 80 \%, 10 \%, and 10 \% of the available data, respectively. The size of the training, validation and test set are referred to by $n_\mathrm{data}^\mathrm{train}$, $n_\mathrm{data}^\mathrm{val}$, and $n_\mathrm{data}^\mathrm{test}$, respectively.

\paragraph{Data normalization}
Min-max normalization is utilized, whereas the minimum and maximum values are taken from the training set to avoid data leakage.

\paragraph{Data augmentation}
The normalized data is augmented to virtually extend the training database and, therewith, setup a more robust ML model \cite{oviedo_fast_2019,li_genetic_2014,choudhury_artificial_2011}. In this work, a generalized version of the constrained mixup augmentation is utilized. Input and output samples are determined by $\hat{x}_{ij}=\lambda x_{i} + (1-\lambda) x_{j}$ and $\hat{y}_{ij}=\lambda y_{i} + (1-\lambda) y_{j}$, respectively \cite{zhang_mixup_2018}. Hence, a hypothesis of linear superposition is provided to the network. Its validity is reflected by the probability distribution function (i.e., Beta($\alpha$)) of the $\lambda$ value. $\alpha$ approaching zero, one, or infinity resembles a coinflip, uniform distribution, or 0.5, respectively. In this work, samples $i$ and $j$ are only mixed up when the length of the vector pointing from one to another falls below $r_\mathrm{c}$, that is $\sqrt{\sum_{k=1}^{n_x}(x_{i}-x_{j})^2}<r_\mathrm{c}$. $n_x$ is the number of input parameters. $r_\mathrm{c}$ is considered a HP. The augmented data set size equals the original training data set size ($n_\mathrm{data,aug}^\mathrm{train}=n_\mathrm{data}^\mathrm{train}$). The training data is augmented anew once per epoch. 

\paragraph{Training procedure and metrics}
Backpropagation of the mean absolute errors (MAEs) is used to update the internal degrees of freedom of the ANNs (e.g., weights) once per batch. The stochastic gradient descent algorithm adaptive moment estimation (Adam) is applied \cite{kingma_adam_2015}. The applied batch size $n_\mathrm{batch}$ is defined to match the set up and the ideal batch size $n_\mathrm{batch,ideal}$ included as HP as close as possible, but required to fulfill $|n_\mathrm{batch}-n_\mathrm{data,aug}^\mathrm{train}\%n_\mathrm{batch}|\leq n_\mathrm{batch,ideal}$. Hence, all data samples contribute almost equally to the learning progress.

The learning rate $r_\mathrm{l}$ is initialized with $r_\mathrm{l\text{-}0}$ and kept constant for a simulated annealing phase, that is outlined later. Afterwards, it is divided by ten whenever the validation MAE falls below its previous minimum value over the course of $n_\mathrm{l\text{-}patience}$ epochs. Early stopping stops the training when there is no further reduction of the validation MAE after $2.5n_\mathrm{l\text{-}patience}$ epochs.

\paragraph{Hyperparameter study}
10-fold Monte Carlo cross validation (MCCV) is utilized to determine a more accurate final validation MAE. Training, validation, and test data sets are randomly selected according to the given split. In 10-fold MCCV, an ensemble of 10 ANNs is trained with 10 different random splits. It used in the selection of the best HPs using an evolution strategy (described later). The coefficient of determination $R^2$ is introduced as an additional, secondary metric. The test set is meant to provide an unbiased measure for the ML model's performance. The last is determined even more thoroughly by applying a 100-fold MCCV for the eventually selected set of HPs to compute the final training, validation and test errors.

\paragraph{Production run}
The test set is not required for the production run. Thus, it is combined with the training set, that is 90 \% of the data. An ensemble of ten ML models is set up and trained to reduce the bias introduced to the model by splitting the data into the two subsets \cite{vanpoucke_small_2020}.

\subsection{Physics-separating artificial neural network methodology}
\label{ssec:model}

\begin{figure}
\includegraphics[width=8cm]{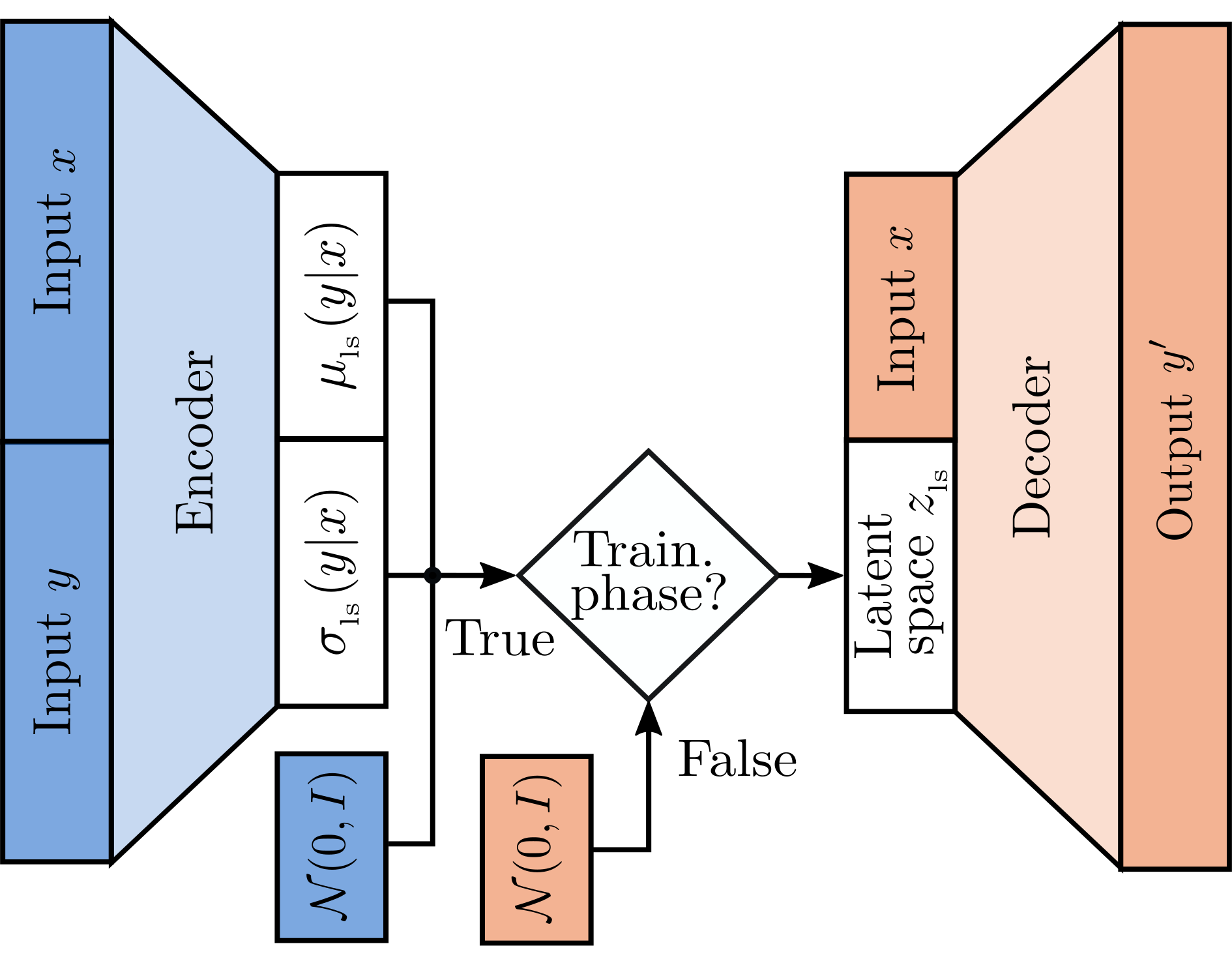}
\caption{Schematic of the CVAE structure. Input variables enter from the left and predictions are extracted on the right of the graph. The coordinates $z_\mathrm{ls}$ of the latent space are indicated by the center white box. The figure is taken from \cite{gergs_physics-separating_2022}.}
\label{fig:CVAE_Schematic}
\end{figure}

\paragraph{Conditional variational autoencoders}
The proposed PSNN combines two regression ML models (i.e., PSI-CVAE, Diffusion-CVAE), that are implemented as conditional variational autoencoders (CVAEs) \cite{gergs_physics-separating_2022,sohn_learning_2015,doersch_tutorial_2021}. Their network architecture and information flow are shown schematically in Figure~\ref{fig:CVAE_Schematic}. CVAEs resemble $\beta$-variational autoencoders ($\beta$-VAEs \cite{kingma_auto-encoding_2013, rezende_stochastic_2014, higgins_beta-vae_2016, burgess_understanding_2017, doersch_tutorial_2021}), whose encoder and decoder are conditioned on the regression input variables. These are set up symmetrically. The number of hidden layer $n_\mathrm{hl}$ and nodes per layer $n_\mathrm{npl}$ are considered as HPs. The activation functions for any hidden and output layer are set as rectified linear unit (ReLU) and linear, respectively. The encoder projects information of the regression output variables $y_i$ to an $n_\mathrm{ls}$ dimensional latent space representation. Similarity to a standard normal distribution in latent space is enforced by introducing an additional Kullback-Leibler (KL) divergence. \cite{kingma_auto-encoding_2013, rezende_stochastic_2014}. The HP $\beta$ is used to scale the KL loss. It is additionally scaled with a simulated annealing factor that is increased logarithmically from 10$^{-3}$ to 1 per batch over the course of $n_\mathrm{SA}$ (also a HP) epochs. The decoder is conditioned on the regression input and the latent space (a standard normal distribution after successful training) tries to reconstruct the regression output. The decoder resembles the regression model to be utilized for prediction (after training is completed). CVAEs are described in detail in \cite{sohn_learning_2015,doersch_tutorial_2021}. The outputs of the CVAEs are passed through physics-constraint enforcing custom layers.

\paragraph{Physics-constraints}
The physics-constraints enforced by the last output layer simplify the regression problem to be solved by the individual CVAEs and are described in the following. First, the suppression of extrapolation is outlined. Second, the particle conversation for prediction on bulk diffusion processes are introduced. Note that predicted quantities are denoted by primes (e.g., $y^\prime$).

\emph{1. Extrapolation Suppression}
Constrained predictions were utilized in previous works to secure physically plausible predictions (e.g., an Ar concentration in the range of 0\,\% to 100\,\%) \cite{gergs_physics-separating_2022}. This procedure is developed further and generalized in this work. In general ML models are well suited for interpolation but often fail to extrapolate beyond known input data. Hence, predictions below (beyond) the minimal (maximal) training reference $y_{\mathrm{min}}$ ($y_{\mathrm{max}}$) are suppressed by folding them back three times to facilitate a more stable system state evolution and guarantee positive quantities when required (e.g., mass density, sputter yield):
\begin{subequations}
\begin{alignat}{6}
y^\prime &\rightarrow 2y_{\mathrm{min}}-y^\prime &\hspace{3cm}&\mathrm{if}&~~& y^\prime &\leq y_{\mathrm{min}}
\label{eq:min_mirror}\\
y^\prime &\rightarrow 2y_{\mathrm{max}}-y^\prime &&\mathrm{if}&& y^\prime &\geq y_{\mathrm{max}}
\label{eq:max_mirror}
\end{alignat}
\end{subequations}

\emph{2. Particle conservation (diffusion)}
The absolute number of Ar, Al, and N atoms must be conserved during bulk diffusion processes, which are modeled by the Diffusion-CVAE. This also demands a balance of the individual point defect populations. Using three corresponding constraints (e.g., based on Eqs.~\eqref{eq:xAr}-\eqref{eq:xAl}) to determine them reduces the number of the ML model's output descriptors, but may eventually contradict the extrapolation suppression constraint introduced in the preceding paragraph. For example, the conservation of Ar atoms prior and post diffusion (prediction) could be realized by determining the Ar population occupying Al lattice sites $\rho_\mathrm{Ar_{Al}}^\prime=n_\mathrm{tot}/n_\mathrm{tot}^\prime(\rho_\mathrm{Ar_{i}}+\rho_\mathrm{Ar_{Al}}+\rho_\mathrm{Ar_{N}})-\rho_\mathrm{Ar_{i}}^\prime-\rho_\mathrm{Ar_{N}}^\prime$ and using Eq.~\eqref{eq:NGes}. However, some predictions may require $n_\mathrm{Ar_{Al}}^\prime$ to be negative (i.e., $n_\mathrm{Ar_{i}}+n_\mathrm{Ar_{Al}}+n_\mathrm{Ar_{N}}<n_\mathrm{Ar_{i}}^\prime+n_\mathrm{Ar_{N}}^\prime$), even though the number of Ar atoms occupying Al lattice sites cannot be negative. Enforcing the constraint outlined in the preceding paragraph may resolve the issue, but being evaluated sequentially again may lead to a violation of particle conversation during diffusion processes. Thus, a more careful point defect balancing is required and introduced in the following.

All Ar point defect population predictions (i.e., $\rho_\mathrm{{Ar}_i}^\prime+\rho_\mathrm{Ar_{Al}}^\prime+\rho_\mathrm{Ar_{N}}^\prime$) are multiplied with a correction factor $f_\mathrm{Ar}$:
\begin{equation}
f_\mathrm{Ar} = \frac{n_\mathrm{tot}}{n_\mathrm{tot}^\prime}\frac{\rho_\mathrm{{Ar}_i}+\rho_\mathrm{{Ar}_{Al}}+\rho_\mathrm{{Ar}_{N}}}{\rho_\mathrm{{Ar}_i}^\prime+\rho_\mathrm{{Ar}_{Al}}^\prime+\rho_\mathrm{{Ar}_{N}}^\prime+10^{-7}}
\label{eq:ArScale}
\end{equation}
The deviation of predicted Al and N atoms prior/post diffusion is defined by $\Delta n_\mathrm{Al}$ and $\Delta n_\mathrm{N}$, respectively:
\begin{subequations}
\begin{align}
\begin{split}
\Delta n_\mathrm{Al} =&n_\mathrm{tot}(\rho_\mathrm{Al_i}+\rho_\mathrm{Al_N}-\rho_\mathrm{v_{Al}}-\rho_\mathrm{N_{Al}}-\rho_\mathrm{(N\text{-}N)_{Al}}-\rho_\mathrm{Ar_{Al}})\\
&-n_\mathrm{tot}^\prime(\rho_\mathrm{Al_i}^\prime+\rho_\mathrm{Al_N}^\prime-\rho_\mathrm{v_{Al}}^\prime-\rho_\mathrm{N_{Al}}^\prime-\rho_\mathrm{(N\text{-}N)_{Al}}^\prime-f_\mathrm{Ar}\rho_\mathrm{Ar_{Al}}^\prime)
\label{eq:DeltaAl}
\end{split}\\
\begin{split}
\Delta n_\mathrm{N} =&n_\mathrm{tot}(\rho_\mathrm{N_i}+2\rho_\mathrm{(N\text{-}N)_i}+\rho_\mathrm{(N\text{-}N)_N}-\rho_\mathrm{v_{N}}+\rho_\mathrm{N_{Al}}+2\rho_\mathrm{(N\text{-}N)_{Al}}-\rho_\mathrm{{Al}_N}-\rho_\mathrm{{Ar}_{N}})\\
&-n_\mathrm{tot}^\prime(\rho_\mathrm{N_i}^\prime+2\rho_\mathrm{(N\text{-}N)_i}^\prime+\rho_\mathrm{(N\text{-}N)_N}^\prime-\rho_\mathrm{v_{N}}^\prime+\rho_\mathrm{N_{Al}}^\prime+2\rho_\mathrm{(N\text{-}N)_{Al}}^\prime-\rho_\mathrm{{Al}_N}^\prime-f_\mathrm{Ar}\rho_\mathrm{{Ar}_{N}}^\prime)
\label{eq:DeltaN}
\end{split}
\end{align}
\end{subequations}
with $n_\mathrm{tot}$ as a function of the point defect population following Eq.~\eqref{eq:NGes}. All defect populations but anti-sites are compensated for the particle balancing using these deviations:
\begin{subequations}
\begin{alignat}{6}
\rho_\mathrm{v_{Al}}^\prime &\rightarrow \frac{n_\mathrm{tot}^\prime\rho_\mathrm{v_{Al}}^\prime-\Delta n_\mathrm{Al}}{n_\mathrm{tot}}&&\mathrm{if}&&\Delta n_\mathrm{Al}&<0
\label{eq:AliV1}\\
\rho_\mathrm{Al_{i}}^\prime &\rightarrow \frac{n_\mathrm{tot}^\prime\rho_\mathrm{Al_{i}}^\prime+\Delta n_\mathrm{Al}}{n_\mathrm{tot}}&&\mathrm{if}&&\Delta n_\mathrm{Al}&>0
\label{eq:AliI1}\\
\rho_\mathrm{v_{N}}^\prime &\rightarrow \frac{n_\mathrm{tot}^\prime\rho_\mathrm{v_{N}}^\prime-\Delta n_\mathrm{N}}{n_\mathrm{tot}}&&\mathrm{if}&&\Delta n_\mathrm{N}&<0
\label{eq:NiV1}\\
\rho_\mathrm{(N\text{-}N)_{N}}^\prime &\rightarrow \frac{n_\mathrm{tot}^\prime\rho_\mathrm{(N\text{-}N)_{N}}^\prime + \frac{\rho_\mathrm{(N\text{-}N)_{N}}^\prime}{\rho_\mathrm{N_i}^\prime + \rho_\mathrm{(N\text{-}N)_{N}}^\prime} \Delta n_\mathrm{N}}{n_\mathrm{tot}}&\hspace{3cm}&\mathrm{if}&~~& \Delta n_\mathrm{N}&>0
\label{eq:NiI1}\\
\rho_\mathrm{N_i}^\prime &\rightarrow \frac{n_\mathrm{tot}^\prime\rho_\mathrm{N_i}^\prime + \frac{\rho_\mathrm{N_i}^\prime}{\rho_\mathrm{N_i}^\prime + \rho_\mathrm{(N\text{-}N)_{N}}^\prime} \Delta n_\mathrm{N}}{n_\mathrm{tot}}&&\mathrm{if}&& \Delta n_\mathrm{N}&>0
\label{eq:NiI2} 
\end{alignat}
\end{subequations}
All point defect populations, which have not been altered up this point, are scaled with the quotient $n_\mathrm{tot}^\prime/n_\mathrm{tot}$ to account for the changed total number of atoms, ensuring consistent predictions.

\begin{figure*}
\includegraphics[width=16cm]{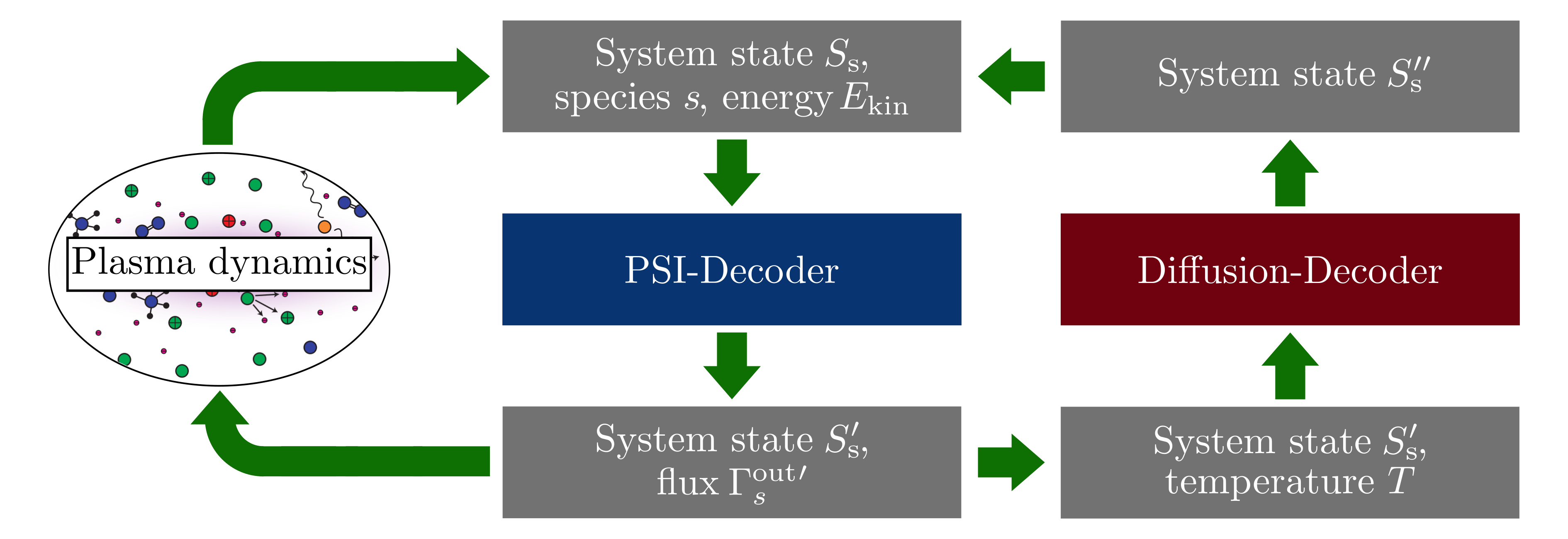}
\caption{Schematic of the PSNN structure and information flow. Plasma dynamics can be imposed by hand, simulation or experiment.}
\label{fig:PSNN}
\end{figure*}

\paragraph{Physics-separating artificial neural network}
Each CVAE (i.e., PSI-CVAE, Diffusion-CVAE) describes one physical process, separating one from another. The (trained) decoders are combined to form a PSNN that allows for an evolution in time by passing the surface state $S_\mathrm{s}$ from one surrogate model to another. The information flow of the PSNN is depicted in Figure~\ref{fig:PSNN}. It resembles closely the information flow inherent to the physical simulations (Fig.~\ref{fig:DataGeneration}. The input to the PSI-Decoder is a single particle sampled from the particle flux of the plasma, characterized by the particles' kinetic energy $E_\mathrm{kin}$, species $s$, and surface state $S_\mathrm{s}$. It predicts the updated surface state $S_\mathrm{s}^\prime$ and emitted flux for each species $\Gamma^{\mathrm{out}\,\prime}_s$. The former is fed together with the temperature $T$ to the Diffusion-Decoder, which predicts a new surface state $S_\mathrm{s}^{\prime\prime}$. It is passed on to the PSI-Decoder, establishing an recurrent link within the PSNN. Note that the direct correspondence of the physical simulations and the separated PSI-Decoder and Diffusion-Decoder structure allows for an efficient parameter space exploration, as outlined in Section~\ref{ssec:RMD}. However, relying on single PSIs, the predictions after training may be subject to vastly different plasma conditions and are not limited to the specific flux ratios or ion energy distributions used for setting up the data set.

\subsection{Hyperparameter study}
\label{ssec:hpStudy}

The HP of each CVAE (i.e., PSI-CVAE, Diffusion-CVAE) are optimized by applying an individual anisotropic self-adaptive evolution strategy with intermediate recombination $(\mu/\mu_\mathrm{I},\lambda)$-$\sigma$SA-ESs. $\mu$, $\lambda$, and $\sigma$ refer to the number of parents, populations size, and step sizes (mutation strengths), respectively. Generalized and topic-wise related descriptions of this method can be found in \cite{back_basic_1994,schwefel_numerische_1977,rozenberg_handbook_2012,gergs_physics-separating_2022}. The HPs considered in this work and their initialization ranges are listed in Table~\ref{table: HP_params}.

The evolution strategies are inialized as $(7/7_\mathrm{I},70)$-$\sigma$SA-ESs. The population sizes $\lambda$ are reduced by one per generation over the course of 63 generations and, afterwards, kept constant. The numbers of parents are determined by $\mu=\lambda/7$ (integer values are enforced) \cite{schwefel1987collective}. The ESs are conducted for 200 generations and, hence, end as $(1,7)$-$\sigma$SA-ESs.

\begin{table}
\caption{The HPs to be optimized, their initialization range, and final values for the PSI-CVAE as well as Diffusion-CVAE.}
\label{table: HP_params}
\begin{center}
\begin{tabular}{ l   c   c   c}
\hline
HP & Init. range & PSI-CVAE & Diffusion-CVAE\\
\hline
$r_\mathrm{c}$		& [$0.0$,$1.0$] & 0.50 & 0.44 \\
$\alpha$		& [$10^{-5}$,$1.0$] & 0.47  $\cdot 10^{-2} $& 0.17 \\
$r_\mathrm{l\text{-}0}$		& [$10^{-3}$,$10^{-2}$] & 9.14  $\cdot 10^{-3}$ & 1.08 $\cdot 10^{-3}$ \\
$n_\mathrm{l\text{-}patience}$& [4,7] & 9 & 7 \\
$\lambda_\mathrm{L2}$	& [0,$10^{-4}$] & 2.23 $\cdot 10^{-7}$ & 1.22 $\cdot 10^{-7}$ \\
$n_\mathrm{hl}$ 		& [1,5] & 1 & 3 \\
$n_\mathrm{npl}$		& [8,128] & 107 & 155 \\
$n_\mathrm{ls}$			& [1,6] & 1 & 5 \\
$\beta$					& [$10^{-1}$,10] & 56.69 & 0.14 \\
$n_\mathrm{SA}$  & [1,$10^{2}$] & $10^{2}$ & $10^{2}$ \\
$n_\mathrm{batch,ideal}$		& [16,64] & 37 & 53\\
\hline
\end{tabular}
\end{center}
\end{table} 

\subsection{Production run: Reference experiment}
\label{ssec:production}

First, the experimental scenario considered for production as well as validation is outlined. Second, the fluxes onto the AlN surfaces required for the ML simulation are calculated and used to introduce an estimated process time. 

\paragraph{Reference experiment}
Ries et al. used a large-area multi-frequency capacitively coupled plasma (MFCCP) to sputter deposit AlN with Ar and N$_2$ as working gases (Ar/N$_2$ gas inlet ratio equal 8/1) \cite{ries_ion_2019}. The electrical asymmetry effect was taken advantage of to decouple the ion flux from the ion energy. The former was kept approximately constant and the latter was controlled by applying voltage waveform tailoring (i.e., adjusting the relative phase shift between the two excitation frequencies). Four cases with mean ion energies $\overline{E_\mathrm{ion}}$ of 47 eV, 53 eV, 57 eV, and 81 eV were considered. The predominant AlN surface orientation was found to vary as function of the mean ion energy: AlN(002) for $\overline{E_\mathrm{ion}}=$47 eV and $\overline{E_\mathrm{ion}}=$53 eV as well as AlN(100) for $\overline{E_\mathrm{ion}}=$57 eV and $\overline{E_\mathrm{ion}}=$81 eV \cite{ries_ion_2019}.

In this work, the species most relevant for PSI (i.e., Al, N$^+$, N$_2^+$, Ar$^+$) are sampled from the experimentally determined fluxes impinging onto the substrate (cf.\ next subsection). The kinetic energy $E_\mathrm{kin}$ of ions and Al neutrals are sampled from measured ion energy distribution functions (IEDFs) (depicted in Figure~11 of \cite{ries_ion_2019}) and from Monte Carlo transport simulations (Al in pure Ar, but assumed invariant), respectively \cite{ries_ion_2019,trieschmann_transport_2015,Trieschmann_2018}. A threshold for the IEDFs is imposed to avoid sampling from noise. Monte Carlo accept-reject sampling (rejection sampling) is used to determine the individual particle energies.

The evolution and response of both monocrystalline systems (i.e., AlN(002), AlN(100)) predicted by the PSNN is to be studied as a function of the mentioned four ion energy distribution functions (IEDFs). Each case starts with ideal, defect free AlN and is run until a steady-state is reached, that is approximately 45 minutes (experimental process time). All cases are re-run 100 times to evaluate their statistics accurately. The final results are averaged over the last minute. Intrinsic stresses are determined as a function of the predicted lattice constants by utilizing the third-order Birch-Murnaghan EOS of the ideal, defect free AlN reference system as proposed and validated in \cite{karimi_aghda_unravelling_2021}. The final stresses and compositions predicted by the PSNN are compared to experimentally measured reference values \cite{ries_ion_2019}. Spurious Fe and O concentrations observed in the experiment are substituted with Al and N concentrations to define a comparable reference for the simulation.

\paragraph{Flux and process time estimations} 
The process time $t_\text{p}$ for $n_\mathrm{pi}$ particle impingements may be estimated by the sum over all reciprocal impingement rates, $t_\text{p} = \sum_{i=1}^{n_\mathrm{pi}} 1/(\Gamma_{i} A_\mathrm{RMD}^\prime)$. $\Gamma_{i}^\mathrm{in}$ and $A_\mathrm{RMD}^\prime$ denote the experimental particle flux onto the AlN surface and predicted RMD AlN surface area, respectively. $A_\mathrm{RMD}^\prime$ is computed as a function of the predicted lattice constant $a^\prime$ and imposed surface orientation. 

The ion flux onto the target and substrate is assumed to be approximately equal for the given geometry and approximated by $\Gamma_\mathrm{ion}^\mathrm{in}=h n_\mathrm{e} v_\mathrm{B}$, with assumed edge-to-center ration $h=0.61$, electron density $n_\mathrm{e}=5\cdot10^{15}~\mathrm{m}^{-3}$, and Bohm velocity $v_\mathrm{B}=3.21 \cdot 10^{3}~\mathrm{m/s}$ (for Ar$^+$ ions and a given electron temperature of $k_\mathrm{B} T_\mathrm{e}=3$ eV) \cite{bienholz2014,lieberman_principles_2005}.  

The flux of Al neutrals onto the substrate is calculated from $\Gamma_\mathrm{Al}^\mathrm{in}= c_\text{t} Y_\mathrm{Ar^+}(352~\mathrm{eV}) \, \Gamma_\mathrm{ion}^\mathrm{in} = 3.47 \cdot 10^{18}~\mathrm{m}^{-2}\mathrm{s}^{-1}$ with the collisional transport coefficient $c_\text{t}=0.6$ obtained from Monte Carlo transport simulations (Al in pure Ar, but assumed invariant) as well as an Ar sputtering yield $Y_\mathrm{Ar^+}(352~\mathrm{eV})=0.579$ (clean Al target) \cite{trieschmann_transport_2015,ries_ion_2019}. The Al flux from the target is obtained by multiplying the ion flux $\Gamma_\mathrm{ion}^\mathrm{in}$ with the Ar$^+$ sputtering yield $Y_\mathrm{Ar^+}$.

The considered ion fluxes (i.e., $\Gamma_\mathrm{N^+}^\mathrm{in}$, $\Gamma_\mathrm{N_2^+}^\mathrm{in}$, $\Gamma_\mathrm{Ar^+}^\mathrm{in}$) are determined by assuming that the composition of the ion fluxes onto the substrate $\Gamma_\mathrm{ion}^\mathrm{in}$ resemble the volumetric composition. The total gas density is given by $n_\mathrm{g,tot}=p/(k_\mathrm{B}T_\mathrm{g})$ with the Boltzmann constant $k_\mathrm{b}$, and gas temperature $T_\mathrm{g}=650~K$ \cite{bienholz2014}. The species specific gas densities are calculated as relative fractions assuming $n_\mathrm{g,N}/n_\mathrm{g,N_2}=1/9$ and $(n_\mathrm{g,N}+n_\mathrm{g,N_2})/n_\mathrm{g,Ar}=1/8$ \cite{ries_ion_2019}. Hence, the working gas approximately consists of 1.16 \% N, 10.47 \% N$_2$, and 88.37 \% Ar. The ion (Bohm) flux onto the substrate is split up accordingly: 
$\Gamma_\mathrm{Al^+}=3.11 \cdot 10^{14}~\mathrm{m}^{-2}\mathrm{s}^{-1}$, $\Gamma_\mathrm{N^+}=1.14 \cdot 10^{17}~\mathrm{m}^{-2}\mathrm{s}^{-1}$, $\Gamma_\mathrm{N_2^+}=1.03 \cdot 10^{18}~\mathrm{m}^{-2}\mathrm{s}^{-1}$, and $\Gamma_\mathrm{Ar^+}=8.65 \cdot 10^{18}~\mathrm{m}^{-2}\mathrm{s}^{-1}$. The contribution due to Al$^+$ is neglected due to their rare occurrence.


\section{Results}
\label{sec:results}

\subsection{Hyperparameter study}
\label{ssec:hpStudy_results}

Following the outlined evolution strategy with MCCV, an optimum set of HPs is determined and listed in Table~\ref{table: HP_params}. As apparent, data augmentation by means of constrained mixup augmentation is beneficial for the Diffusion-CVAE ($\alpha=0.17$). This means that the hypothesis of linear superposition is accepted to some extend for the diffusion processes but declined for the PSIs ($\alpha=0.47 \cdot 10^{-2}$). Kernel regularization is found to be disadvantageous for either ML model ($\lambda_\mathrm{L2} \approx 10^{-7}$). The network structure of the Diffusion-CVAE (i.e., 3 hidden layer with 155 nodes per layer) allows for higher order of complexity than the PSI-CVAE's one (i.e., 1 hidden layer with 107 nodes per layer). It is also interesting to note that the optimum number of simulated annealing epochs is 100 for both ML models, which is the imposed upper boundary for this HP for the evolution strategies. Hence, the simulated annealing step is assumed to be of great use for the training procedure.

In addition to the MAE, the performance of the PSI-CVAE and Diffusion-CVAE with their final set of HPs listed in Table~\ref{table: HP_params} can be assessed by the coefficient of determination $R^2$. It is calculated on the training, validation, and test set to equal 0.87, 0.86, and 0.87 for the PSI-CVAE as well as 0.94, 0.93, and 0.93 for the Diffusion-CVAE, respectively. These values $\gtrsim 0.9$ signify an accurate model approach ($R^2=1$ signifies fully explained variance in the data). The negligible difference between the three subsets indicates that the ML models learned successfully to generalize on the training data. This finding is analyzed more thoroughly in the following by comparing the unnormalized mean absolute errors (MAEs) of each system property (e.g., mass density). It is important to note though that the reference data does not resemble any kind of ground truth but contains statistical fluctuations (e.g., a single ion hitting the surface on a different surface sites is likely to inflict different kinds of defect structures) which intrinsically provide limits for the MAEs.

\begin{figure*}
\includegraphics[width=16cm]{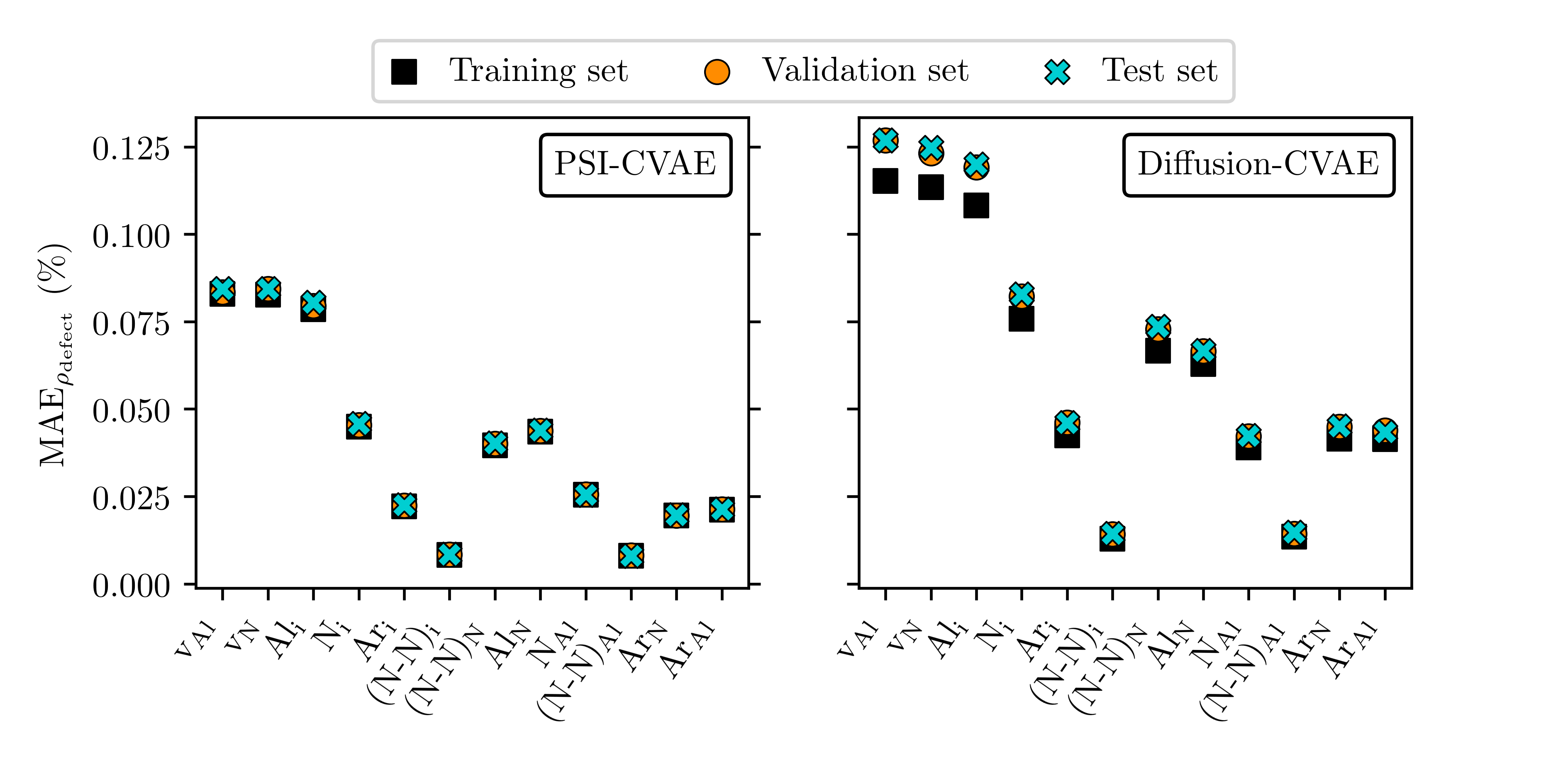}
\caption{MAE of the unnormalized predictions on the point defect populations and data sets. Point defect types are listed on the $x$-axis.}
\label{fig:defect_population_HP}
\end{figure*}

The MAE of all considered defect populations are shown in Figure~\ref{fig:defect_population_HP}. The PSI-CVAE and Diffusion-CVAE is found to predict the defect structure accurately with errors that are of the order/below 0.1 \%. The error of the PSI-CVAE's predictions on the training, validation, and test set are barely distinguishable from each other, resembling excellent generalization. The Diffusion-CVAE is found to perform best on the training set, showing minor signs of overfitting. However, the difference between the validation and test set is negligible.

\begin{table}
\caption{MAE of the unnormalized predictions on all data sets for the PSI-PSNN.}
\label{table: PSIPSNN_HP}
\begin{center}
\begin{tabular}{ l   c   c c}
\hline
Property & Train. set & Val. set & Test set \\
\hline
$a$ (\r A)									& 0.002 & 0.002 & 0.002 \\
$\Delta E_\mathrm{f}$ (eV)					& 0.005 & 0.005 & 0.005 \\
B (GPa)										& 3.029 & 3.067 & 3.055 \\
B$^\prime$ (GPa)							& 0.913 & 0.922 & 0.931 \\
$\Gamma_\mathrm{Al}^\mathrm{out}/\Gamma^\mathrm{in}_s$ (.)	& 0.015 & 0.016 & 0.015\\
$\Gamma_\mathrm{N}^\mathrm{out}/\Gamma^\mathrm{in}_s$ (.)		& 0.089 & 0.089 & 0.087\\
$\Gamma_\mathrm{Ar}^\mathrm{out}/\Gamma^\mathrm{in}_s$ (.)	& 0.218 & 0.219  & 0.219 \\
$\Gamma_\mathrm{N_2}^\mathrm{out}/\Gamma^\mathrm{in}_s$ (.)	& 0.139 & 0.139 & 0.140 \\
\hline
\end{tabular}
\end{center}
\end{table}

\begin{table}
\caption{MAE of the unnormalized predictions on all data sets for the Diffusion-PSNN.}
\label{table: DiffusionPSNN_HP}
\begin{center}
\begin{tabular}{ l   c   c c}
\hline
Property & Train. set & Val. set & Test set \\
\hline
$a$ (\r A)									& 0.001 & 0.001 & 0.001 \\
$\Delta E_\mathrm{f}$ (eV)					& 0.007 & 0.008 & 0.008 \\
B (GPa)										& 2.905 & 3.094 & 3.124 \\
B$^\prime$ (GPa)							& 0.918 & 0.963 & 0.973 \\
\hline
\end{tabular}
\end{center}
\end{table}

The high accuracy prediction of the PSI-CVAE and the Diffusion-CVAE on the lattice constant $a$, the formation energy $\Delta E_\mathrm{f}$, the bulk modulus B, and its derivative B$^\prime$ are presented in Table~\ref{table: PSIPSNN_HP} and Table~\ref{table: DiffusionPSNN_HP}, respectively. The almost interchangeable performance on the training, validation, and test set shows again that the models successfully learned to generalize on the provided data. However, the MAEs of the emitted Al, N, Ar and N$_2$ flux per incident flux, as listed in Table~\ref{table: PSIPSNN_HP}, are relatively large when compared to typical sputter yields as well as reflection ratios in the considered regime of kinetic energies (i.e., $E_\mathrm{kin}$ in [0  eV, 300 eV]). It is argued that these larger errors do not signify bad performance, but are rather a consequence of the data assembly for the PSIs. One PSI data sample contains the information on a single PSI, which leads to the emission of, for example, none, one, or maybe two particles. This will be perceived as noise to the ML model, which consequently learns to predict the mean number of emitted particles per PSI for a given surface state. This inherently leads to relatively large MAEs but ultimately is exactly what the PSI-CVAE is meant to learn.

\subsection{Production run}
\label{ssec:production_results}

The production run resembles the reference experiment of AlN thin-film deposition for four discharge conditions as previously discussed. In the following, they are investigated for two surface orientations (100) and (002). Initially the emitted particle fluxes are discussed:
Particles are emitted from the surface due to reflection of the incident particle or sputtering of surface atoms. It is observed that most fluxes reach a steady-state after a few seconds. Minor changes on the minute time-scale are observed only for three cases (i.e., $\overline{E_\mathrm{ion}}=$47 eV: (002), $\overline{E_\mathrm{ion}}=$53 eV: (002), $\overline{E_\mathrm{ion}}=$57 eV: (100)) due to a change of the Al sticking probability of approximately 0.5 \%. This transient variation is a side-effect of slowly evolving system states, described in detail later. All Al sticking coefficients are in between 98-99 \%.

\begin{figure*}
\includegraphics[width=16cm]{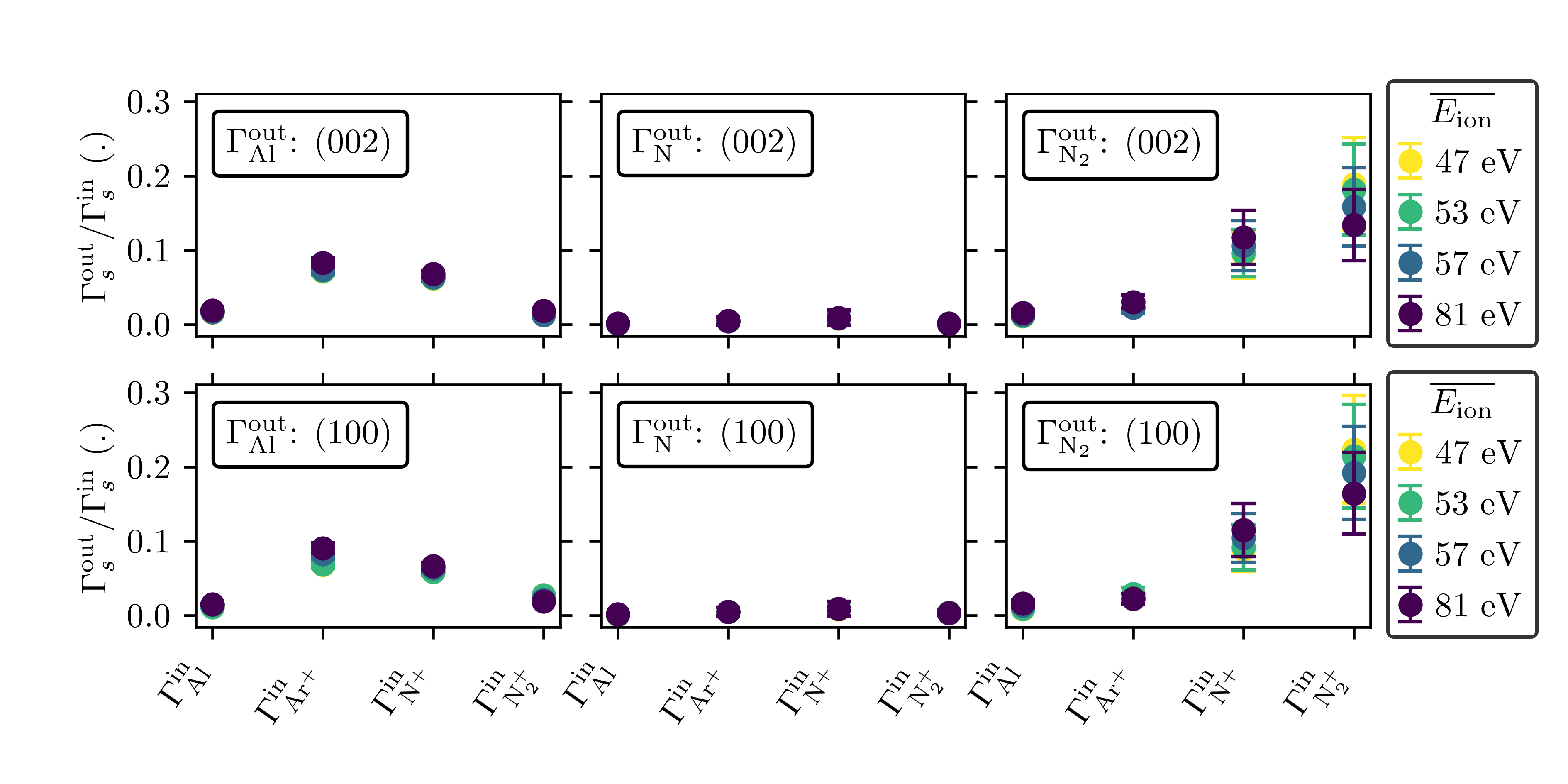}
\caption{The emission of all film forming species per incident fluxes are presented for all considered IEDFs as well as surface orientations. Circle and error bars represent mean values and root-mean-squared deviations, respectively.}
\label{fig:Fluxes}
\end{figure*}

The emitted per incident particle fluxes averaged over the last, 45th minute are shown in Figure~\ref{fig:Fluxes} for all film forming flux combinations (i.e., the emission of Ar is omitted). No significant difference between the two surface orientations is recognizable, which is attributed to the considered ion energy regime of 30 to 100 eV. Higher ion energies are expected to present surface orientation dependent sputtering yields. 

The impingement of N$^+$ and Ar$^+$ ions leads to an almost similar removal of Al atoms, whereas Ar$^+$ ions achieve a slightly increased Al sputtering yield (i.e., $\Gamma_\mathrm{Al}^\mathrm{out}/\Gamma_\mathrm{N^+}^\mathrm{in}<\Gamma_\mathrm{Al}^\mathrm{out}/\Gamma_\mathrm{Ar^+}^\mathrm{in}$). This is attributed to elastic collisions of bombarding N$^+$ ions with N surface atoms, distributing the momentum more rapidly and evenly among them than Al atom. The displacement of N atoms in the subsurface regions leads to the temporary formation of (N-N)$_\mathrm{N}$ close to the surface, where they eventually leave as N$_2$. Higher ion energies lead to deeper collision cascades spawned with higher momenta. The proportionality with the mean ion energy indicates that for neither IEDF a relevant proportion of N$^+$ ions directly form temporary (N-N)$_\mathrm{N}$ at the surface (and desorb as N$_2$). 

Bombarding N$_2^+$ ions are split apart when they hit the surface and, thus, inhibit a reduced individual momentum compared to the initially shared one. This favors an even stronger distribution of the momenta in the surface slab and, thus, lessens the likelihood of sputtering Al atoms in the considered ion energy regime. Moreover, for smaller ion energies a shallower subsurface region is affected, which enables incident N$_2^+$ ions to directly form temporary (N-N)$_\mathrm{N}$ at the surface before leaving as N$_2$. This is reflected by the decreased flux ratio $\Gamma_\mathrm{N_2}^\mathrm{out}/\Gamma_\mathrm{N_2^+}^\mathrm{in}$ for increased ion energies. 

\begin{figure*}
\includegraphics[width=16cm]{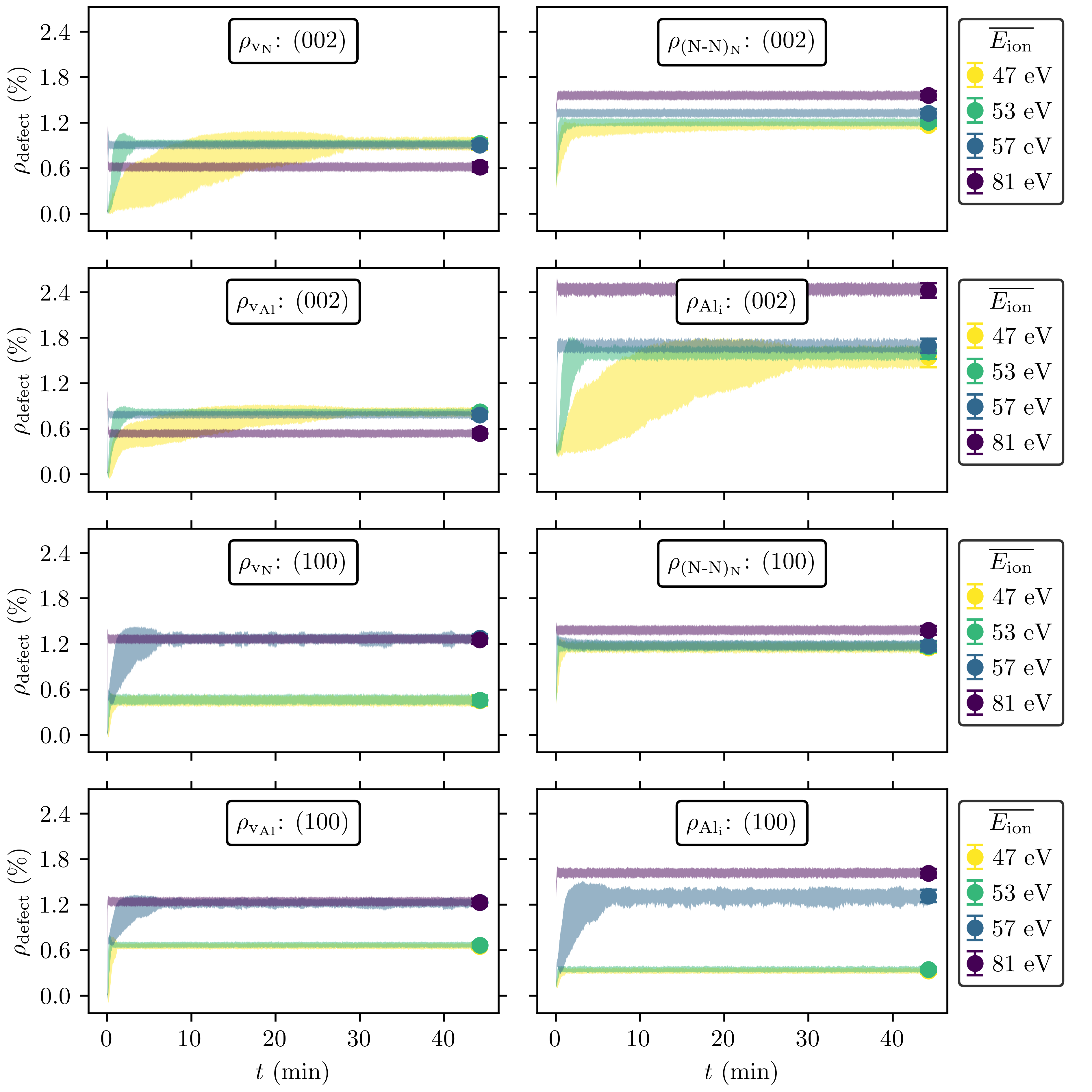}
\caption{Transient evolution of the most relevant point defect populations for all considered IEDFs as well as surface orientations. Error bars and the height of transparent region resemble the mean plus / minus the RMSD.}
\label{fig:defects}
\end{figure*}

The transient evolutions of the most relevant point defect populations are shown in Figure~\ref{fig:defects} for all considered IEDFs and surface orientations. The deposition onto AlN(002) with $\overline{E_\mathrm{ion}}=$47 eV takes up to 30 minutes to reach a steady-state. The ongoing ion bombardment spawns collision cascades in the subsurface region, which once they have worn off may leave vacancies and interstitials behind. Sputtering events or the desorption of N$_2$ remove atoms from the surface and, thus, facilitate the accumulation of vacancies. The Al and N vacancy populations are approximately equal. The Al interstitial population is greater than the N split interstitial population, and both exceed the corresponding vacancy populations. This is due forward sputtering (peening) of surface atoms as well as incorporation of energetic particles (i.e., N$^+$, N$_2$, small proportion of Al), which eventually either reside as interstitials or recombine with vacancies. IEDFs with slightly higher mean ion energies (i.e., 53 eV, 57 eV) converge to a similar point defect structure with marginally increased N split and Al interstitial populations, but require significantly less time for equilibration. These require a few minutes and seconds for  $\overline{E_\mathrm{ion}}=$53 eV and $\overline{E_\mathrm{ion}}=$57 eV, respectively. Therefore it is assumed that for $\overline{E_\mathrm{ion}}=$47 eV scarcely sampled ions with relatively high kinetic energies push the systems to their final state. The likelihood for encountering such ions is naturally increased when increasing the mean ion energy. This effect is enhanced by a change of the IEDF shapes (i.e., narrow unimodal $\rightarrow$ narrow bimodal $\rightarrow$ broad unimodal). A more significantly increased mean ion energy of 81 eV leads to the evolution to a different system state with less Al and N vacancies ($\rho_\mathrm{v_{Al}} \approx \rho_\mathrm{v_{N}}$) and more interstitials ($\rho_\mathrm{{Al}_i} > \rho_\mathrm{(N\text{-}N)_{N}}$). The evolution of the vacancy populations inhibits intermediate maxima after a few seconds. Subsequently, vacancies are removed due to recombination as described before and reach a steady-state after 10 seconds. The evolution of the point defect structures are depicted in Figure~\ref{fig:defects} for up to 45 minutes (and are available in the appendix for up to 100 seconds).

The deposition onto AlN(100) leads to similar system dynamics for $\overline{E_\mathrm{ion}}=$47 eV and $\overline{E_\mathrm{ion}}=$53 eV. The Al and N vacancy populations are approximately equal too. But a greater number of N split and smaller number of Al interstitials are observed. Scarce Al atoms hitting the surface with relative high kinetic energies of up to 30 eV provide an insufficient momentum when penetrating the AlN(100) surfaces to be persistently incorporated, i.e. they end up atop the surface. Increasing the mean ion energy to 57 eV leads to a system evolution that requires up to 10 minutes to reach a steady-state that differs significantly from the previous one. The equilibration on the minute-time scale is again attributed to the contribution of only a small proportion of the incident ions with sufficient kinetic energies, which are pushing the systems to their final state (cf. $\overline{E_\mathrm{ion}}=$53 eV: (002)). The (N-N)$_\mathrm{N}$ populations remain unchanged. The Al and N vacancy populations are doubled. Hence, the probability for the recombination of surface near Al vacancies and incident Al atoms is increased too. The final Al interstitial population are therefore even more than doubled. The point defect structure is characterized predominantly by Al and N Frenkel pairs (vacancies plus interstitials). The evolution to this new system state is caused by a change of the IEDF shapes. The IEDF with $\overline{E_\mathrm{ion}}=$53 eV (narrow bimodal IEDF) and $\overline{E_\mathrm{ion}}=$57 eV (broad unimodal IEDF) reaches up to 60 eV and 75 eV, respectively. The cases with the highest mean ion energies of 83 eV converge to a similar system state with slightly increased interstitial populations, but it takes only a few seconds.

\begin{figure}
\includegraphics[width=8cm]{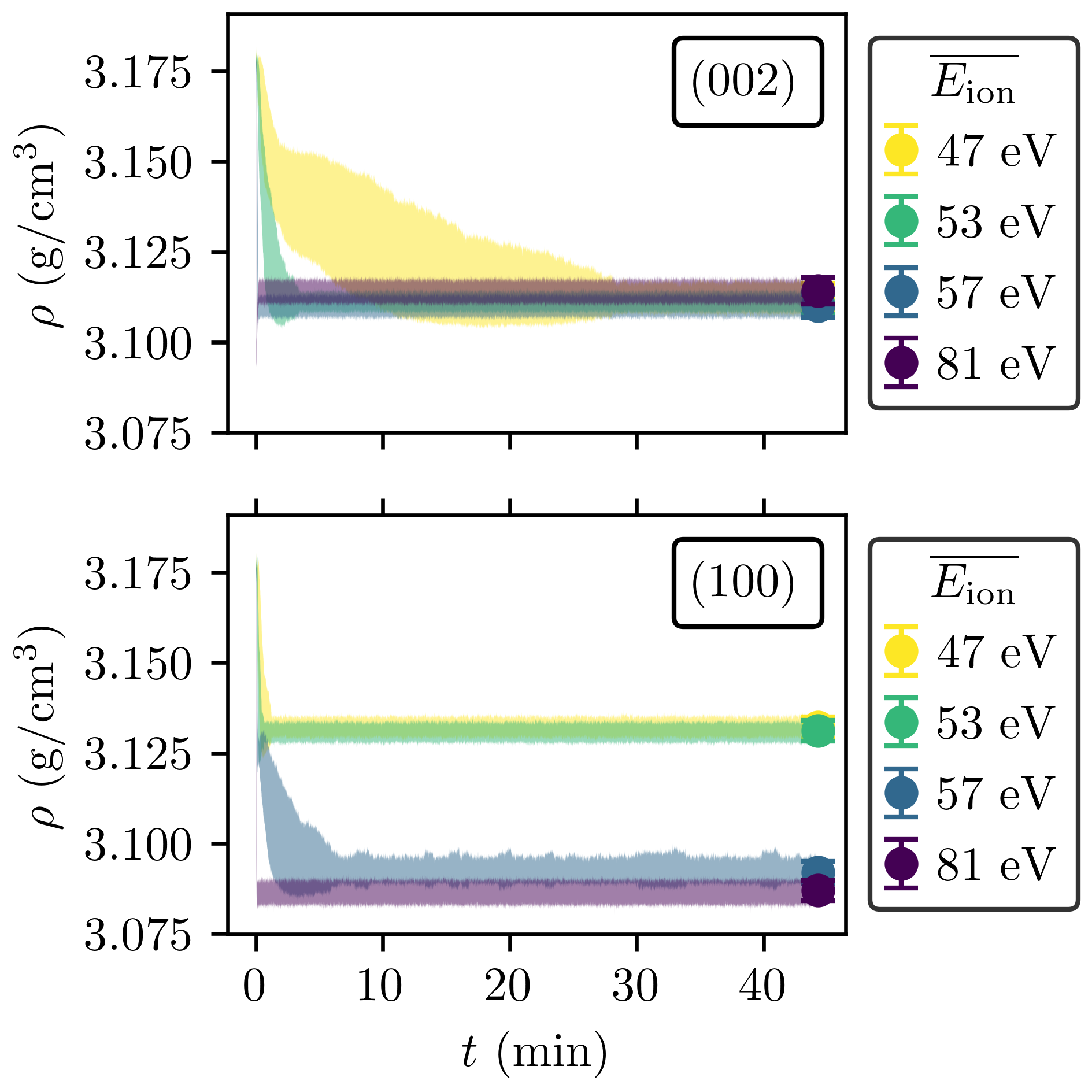}
\caption{Transient evolution of the mass density for all considered IEDFs as well as surface orientations. Error bars and the height of transparent region resemble the mean plus / minus the root-mean-squared deviations.}
\label{fig:density}
\end{figure}

The evolution of the mass densities are presented in Figure~\ref{fig:density}. The equilibration time of the individual cases shows a consistent behavior. However, it is interesting to note that all cases converge to a similar mass density for the surface orientation (002). The accumulation of interstitials is balanced out by a corresponding volumetric expansion. In case of AlN(100), two final point defect structures were discussed in the preceding paragraph. These two system states are reflected by two distinctly separate mass densities. Higher ion energies lead to a great number of Al as well as N Frenkel pairs, which do not alter the mass of the atomic system but cause stress and correspondingly a volumetric relaxation. Hence, smaller mass densities are observed. 

\begin{figure}
\includegraphics[width=8cm]{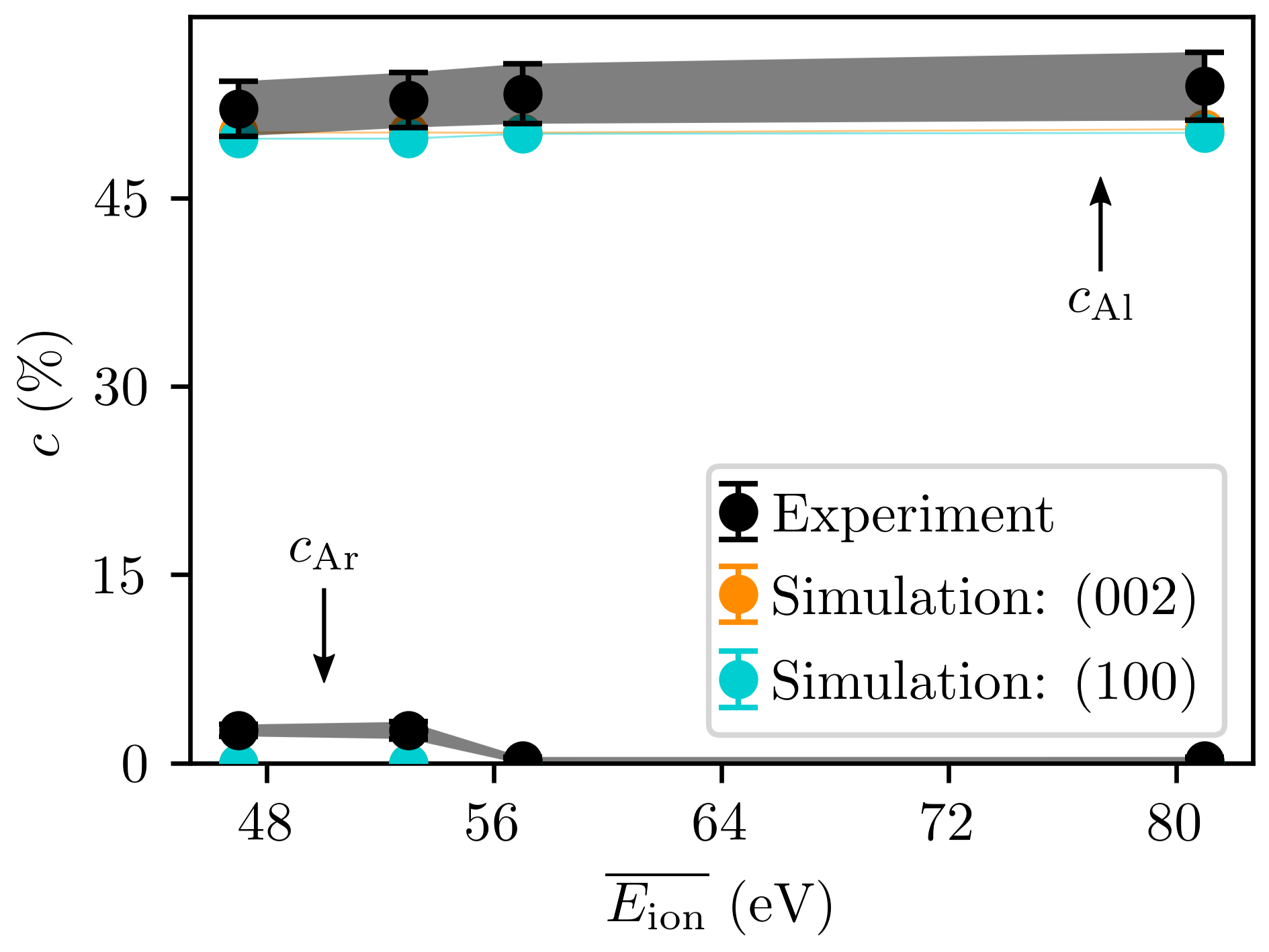}
\caption{The final composition (i.e., Al and Ar concentration $c_\mathrm{Al}$ and $c_\mathrm{Ar}$, respectively) for all considered IEDFs as well as surface orientations averaged over the last minute are compared to experimental reference values \cite{ries_ion_2019}. Circles and error bars represent mean values and root-mean-squared deviations.}
\label{fig:stoichios}
\end{figure}

The composition of the deposited AlN(002) and AlN(100) thin films averaged over the last minute are shown in Figure~\ref{fig:stoichios} in comparison to experimental reference values \cite{ries_ion_2019}. A good agreement with the experiment is achieved when predicting stoichiometric AlN thin films even though the Ar concentration of $2.5\pm0.1$ \% for $\overline{E_\mathrm{ion}}=47$ eV and $\overline{E_\mathrm{ion}}=53$ eV is not reproduced.

\begin{figure}
\includegraphics[width=8cm]{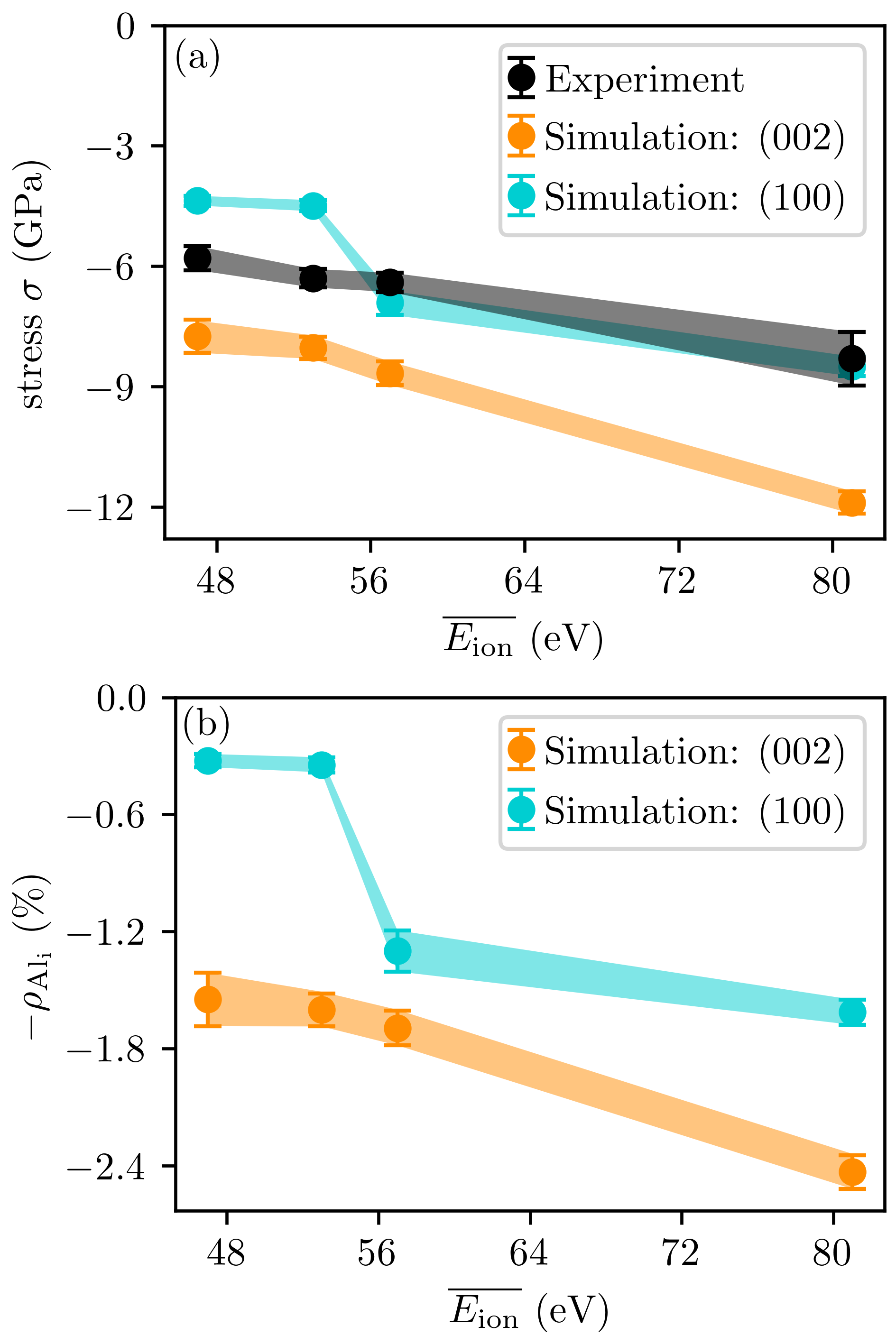}
\caption{The (a) final stress and (b) negated Al interstitial population for all considered IEDFs as well as surface orientations averaged over the last minute are compared to experimental reference values \cite{ries_ion_2019}. Circles and error bars represent mean values and root-mean-squared deviations.}
\label{fig:stresses}
\end{figure}

The stresses predicted by the PSNN and measured in the experiment are presented in Figure~\ref{fig:stresses} (a). An increasingly compressive stress is observed for greater mean ion energies in either case due to the enhanced ion bombardment induced point defect formation. Vacancies and interstitials cause tensile and compressive stresses, respectively. The interplay of all point defects define the film stresses in the ML simulation. However, the contributions due to Al interstitials dominate the stress formation due to their larger size and high formation energies \cite{stampfl_theoretical_2002}. This finding is illustrated by a similar dependence of the stresses and the negated Al interstitial populations (multiplied by -1) on the mean ion energy $\overline{E_\mathrm{ion}}$, as shown in Figures~\ref{fig:stresses}. The preferential surface orientation was found to change from (002) to (100) in the experiment when increasing the mean ion energies from 47-53 eV to 57-81 eV \cite{ries_ion_2019}. By comparison with the ML prediction for the (002) surface orientation, it can be inferred that the predicted stresses for the two IEDFs with smaller mean ion energies (i.e., 47 eV, 53 eV) are overestimated. However, from comparison with the prediction for the (100) surface orientation, the two IEDFs with greater mean ion energies (i.e., 57 eV, 81 eV) are in excellent agreement with the experiment. The change of the predominant surface orientation (i.e., (002)$\rightarrow$(100)) observed in the experiment may be attributed to the reduced compressive stresses predicted to reach up to -12 GPa for (002), compared to -8 GPa for (100).

\section{Conclusion}
\label{sec:conclusion}

This work is meant to further advance the development of data-driven plasma-surface interaction models with atomic fidelity \cite{gergs_physics-separating_2022}. Reactive processes (i.e., sputtering and deposition of AlN in an Ar/N$_2$ discharges) are taken into account. A data-generating scheme is proposed that overcomes the burden of computationally too demanding simulations (i.e., hybrid RMD/tfMC) and, hence, undersampled parameter spaces. The latter are effectively populated by evolving randomly sampled system states $S_\mathrm{s}$ by means of random PSIs (i.e., species $s$, kinetic energy $E_\mathrm{kin}$) and diffusion processes (i.e., temperature $T$). The effect of a single PSI on the deposited film is estimated by cleaving and reinforcing the corresponding bulk structure in surface normal direction \cite{karimi_aghda_unravelling_2021}. A PSNN is used to separate the PSIs from the diffusion processes, which allows for a more efficient data-generation and enforcement of physics-constraints (e.g., particle conservation during bulk diffusion). 

The trained PSNN model is applied to an experimental reference sputter deposition of AlN by taking the corresponding particle fluxes and IEDFs with mean ion energies in the range of 47-81 eV into account \cite{ries_ion_2019}. Ar$^+$ ions are found to remove more Al than N atoms from the surface. The inverse is observed for N$^+$ ions, which spawn collision cascades that distribute their momenta more rapidly with the N surface atoms. This facilitates the temporary formation of (N-N)$_\mathrm{N}$ at the very surface that eventually leave as N$_2$. N$_2^+$ ions are split up when they hit the surface and, thus, spawn two collision cascades with reduced individual momenta compared to the initially shared one. A diminishing amount of Al atoms is sputtered and a shallower subsurface region is effected. The latter allows for the direct formation of (N-N)$_\mathrm{N}$ at the surface and subsequent emission as N$_2$. Atomic nitrogen is rarely sputtered by either ion species. Higher mean ion energies decrease the outgoing flux of N$_2$ due N$_2^+$ ion bombardment but increase the formation of persistent, deeper (N-N)$_\mathrm{N}$. The predicted film depositions take either a few seconds or up to 30 minutes to reach their respective steady-state. Long equilibration times are observed when rare ions whose kinetic energy originates from the high energy tail of the IEDF push the systems to their final states. The latter is found to be dependent on the imposed surface orientation. In particular, a greater Al interstitials population is predicted for AlN(002) than for AlN(100). This point defect type predominantly determines the compressive stress evolution in the deposited AlN thin films. The stresses predicted by the PSNN are quantitatively and qualitatively in good agreement with the experimental reference values in spite of neglecting for instance thermal stresses or point defect annihilation at grain boundaries. The ML model predicts stoichiometric AlN that is observed in the experiment too. 

In summary, 200 million plasma-surface interactions and diffusion processes were predicted with high physical fidelity (hybrid RMD/tfMC). This enabled the evolution of 800 AlN systems (100 $\times$ four IEDFs $\times$ two surface orientations) in time for up to 45 minutes. It took about 34 hours to perform all machine learning predictions with a single GPU. Hence, predictions can be readily extended to cover up the total experimental deposition time of up to hours when required. In contrast, conducting the same case study with hybrid RMD/tfMC simulations is unattainable as it would take more than approximately 8 million CPU years.

\section*{Acknowledgement}

Funded by the Deutsche Forschungsgemeinschaft (DFG, German Research Foundation) -- Project-ID 138690629 -- TRR 87 and -- Project-ID 434434223 -- SFB 1461. The authors thank Dr.-Ing.\ S. Ries from Ruhr University Bochum, S. Karimi Aghda, M. Sc. from RWTH Aachen University, and L. Vialetto, Ph.D. from Kiel University for fruitful discussions.

\section*{Data Availability}

The data that support the findings of this study are available from the corresponding author upon reasonable request.


\section*{ORCID}
\noindent
T. Gergs: \url{https://orcid.org/0000-0001-5041-2941} \\
T. Mussenbrock: \url{https://orcid.org/0000-0001-6445-4990} \\
J. Trieschmann: \url{https://orcid.org/0000-0001-9136-8019}


\clearpage
\appendix
\section*{Appendix}

\begin{figure*}[ht]
\includegraphics[width=16cm]{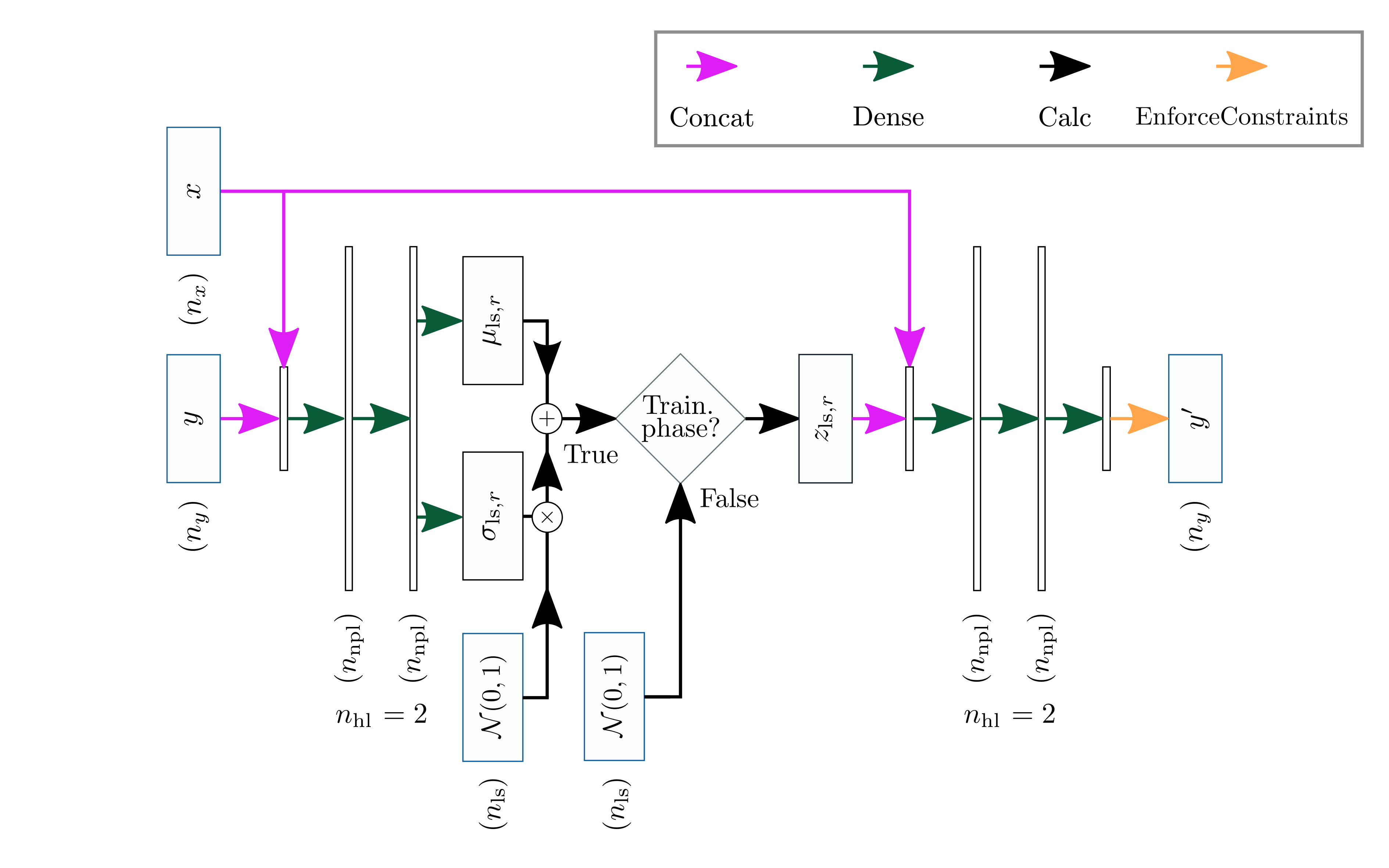}
\caption{Schematic of the CVAE network structure. The shape of the data is provided in parenthesis. Machine learning operations are indicated by colored arrows. The inputs and outputs for the PSI-CVAE are given by $x=\{E_\mathrm{kin},~s,~S_\mathrm{s}\}$ and $y=\{\Gamma_s^\mathrm{out},~S_\mathrm{s}\}$, respectively. The inputs and outputs for the Diffusion-CVAE are given by $x=\{T,~S_\mathrm{s}\}$ and $y=\{S_\mathrm{s}\}$, respectively. }
\label{fig:CVAE_detailed}
\end{figure*}

\begin{figure*}
\includegraphics[width=16cm]{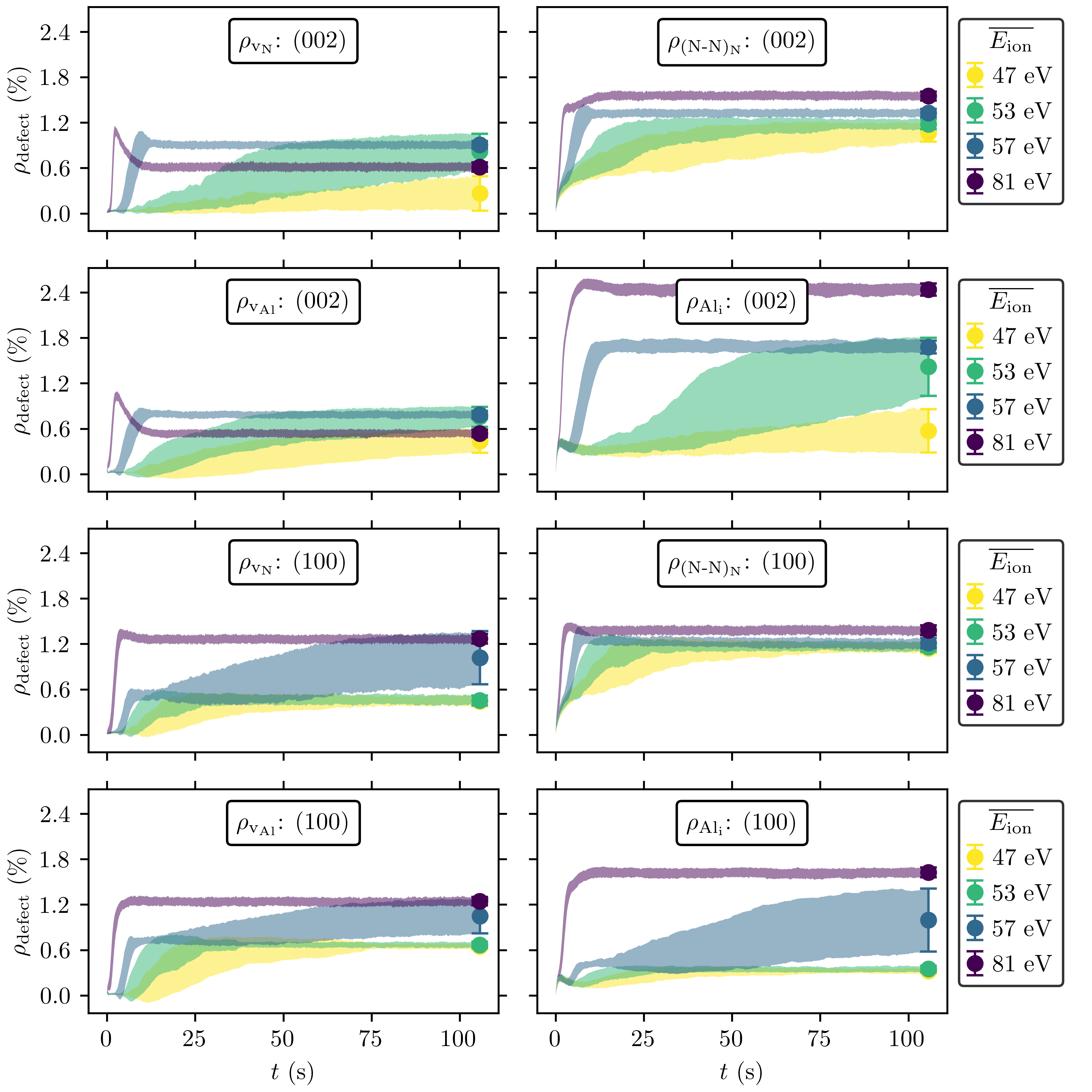}
\caption{Transient evolution of the most relevant point defect populations for all considered IEDFs as well as surface orientations. Error bars and the height of transparent region resemble the mean plus / minus the root-mean-squared deviations.}
\label{fig:defects_short}
\end{figure*}

\clearpage
\bibliography{./references.bib}

\end{document}